\documentclass[amsmath,aps,nofootinbib,superscriptaddress,letterpaper,12pt]{revtex4-2}

\usepackage{color}
\usepackage{amssymb}
\usepackage{amsmath}
\usepackage{epsf}
\usepackage{subfigure}
\usepackage{mathrsfs}
\usepackage{accents}
\usepackage{cancel}
\usepackage{longtable}
\usepackage{mathtools}
\usepackage{relsize}
\usepackage{enumerate}
\usepackage{xfrac}
\usepackage{physics}
\usepackage{bm}
\usepackage{adjustbox}
\usepackage{esint}

\DeclareMathAlphabet{\mathpzc}{OT1}{pzc}{m}{it}
\usepackage{footmisc}

\begin{document}

\title{Great Inequality of Jupiter and Saturn I: The Planetary Three Body Problem, Heliocentric development by Lagrange multipliers, Perturbation Theory Formulation}

\author{
	\textsc{Jonathan Tot} \\
	Department of Mathematics and Statistics, Dalhousie University, \\
	Halifax, Nova Scotia, Canada B3H 4R2 \\
	\normalsize \href{mailto:jonathan.tot@dal.ca}{jonathan.tot@dal.ca}\\
	\,\\
	\textsc{S.R. Valluri}\\
	Department of Physics and Astronomy, University of Western Ontario\\
	\textit{and} Mathematics, King’s University College \\
	London, Ontario, Canada N6A 3K7 \\ 
	\normalsize \href{mailto:valluri@uwo.ca}{valluri@uwo.ca}\\
	\,\\
	\textsc{P.C. Deshmukh} \\
	CAMOST, IIT Tirupati and IISER Tirupati, \\ 
	Tirupati, Andhra Pradesh 517619, India \\ 
	\normalsize \href{mailto:pcd@iittp.ac.in}{pcd@iittp.ac.in}
}
\begin{abstract}
	In this paper, we undertake to present a self-contained and thorough analysis of the gravitational three body problem, with anticipated application to the Great Inequality of Jupiter and Saturn.  The analysis of the three body Lagrangian is very convenient in heliocentric coordinates with Lagrange multipliers, the coordinates being the vector-sides $\vec{r}_i,\,i=1,2,3$ of the triangle that the bodies form.  In two dimensions to begin with, the equations of motion are formulated into a dynamical system for the polar angles $\theta_i$, angular momenta $\ell_i$ and eccentricity vectors $\vec{e}_i$.  The dynamical system is simplified considerably by change of variables to certain auxiliary vector $\vec{f}_i=\hat{r}_i+\vec{e}_i$.  We then begin to formulate the Hamiltonian perturbation theory of the problem, now in three dimensions.  We first give the geometric definitions for the Delaunay action-angle variables of the two body problem.  We express the three body Hamiltonian in terms of Delaunay variables in each sector $i=1,2,3$, revealing that it is a nearly integrable Hamiltonian.  We then present the KAM theory perturbative approach that will be followed in future work, including the modification that will be required because the Hamiltonian is degenerate.
\end{abstract}

\maketitle

\part*{}

\section{Introduction}
The three body-problem and the Great Inequality of Jupiter and Saturn are much studied from the 18th and 19th centuries.  They are part of the history of the development of Celestial Mechanics, inaugurated by the publication of Newton’s Principia in 1687 \cite{principia}, and many of the great names that come down to us from that time were heavily involved, most chiefly the mathematical techniques proposed and analyses completed by Leonhard Euler, Joseph-Louis Lagrange, and finally Pierre-Simon Laplace, in his Théorie de Jupiter et de Saturne, written in 1785 \cite{laplace}.  These advances were enabled by increasing observational accuracy during the time of John Flamsteed, the first Astronomer Royal, and astronomical forecasts of such figures as Cassini and Halley \cite[p.16]{wilson}.

Following the grand realization, due to Laplace, that the observed discrepancies of Jupiter and Saturn’s motions from the essentially Keplerian predictions could actually be accounted for by the mutual Newtonian gravitation of the planets, other researchers into the 19th century continued to make improved theories and contributions to the methods, perhaps most notably the work of G.W. Hill \cite{hill1} and the series manipulations of Charles Delaunay \cite{hill2}.

The study of the three body problem, as a mathematical problem in its own right, also began with the widespread acknowledgement of Newtonian universal gravitation.  Euler (1767) and Lagrange (1772) both initially found particular periodic solutions, which today are understood to be associated with central configurations of the system \cite{quarles}.  Much of the early focus was given to the restricted three body problem, in which one mass is negligible relative to the other two, so that the larger masses form a Keplerian system, and the third body orbits in the gravitational field of the first two.  It was Euler who first set the (circular) restricted three body problem in rotating coordinates, and Lagrange who demonstrated the existence of equilibrium points within the rotating frame, today known as the five Lagrange points, which correspond to five orbits for the third mass with the same orbital frequency as the two-body system.

In the late 1800s, Poincaré studied the general three body problem.  Famously, his work won the prize competition for the 60th birthday of Oscar II, the King of Sweden and Norway in 1885. Poincaré’s work on the three body problem led him to consider what are now known as Poincaré sections and first return maps, and these insights ultimately led to the development of Kolmogorov-Arnold-Moser theory in the mid-20th century \cite{kam}.  In the large, three-volume Les Méthodes Nouvelles de la Mécanique Céleste (1892-99) \cite{poincare}, Poincaré saw, in the unpredictable nature of three-body problem solutions, the first glimpses of chaotic dynamics, which dominates much of dynamical systems analysis today \cite{chaos}.

In this work we present a self-contained analysis of the planetary three body problem, in which two lighter masses orbit a heavier central body, with particular application to the Sun-Jupiter-Saturn system.  Working with modern mathematical notation and physical terminology, one aim of this work is pedagogical in nature, that the subject matter would be more accessible to modern audiences.  For readers who are new to the subject, we put classical terminology in \textit{italics} upon their first occurrence and definition in the text.  Jupiter and Saturn’s conjunction in December 2020 gained much attention in the media \cite{news1,news1}, so popular presentation of this work should serve to foster public engagement in mathematics and the sciences.

Crucial to our analysis is the work of Brouke and Lass \cite{brouke_lass}, who formulate the Lagrangian problem in terms of the three vector-sides of the triangle that the bodies form, using Lagrange multipliers.  In Part~\ref{part1}, we present this treatment of the three body problem in heliocentric coordinates. In \S\ref{constants} we show how total energy and total angular momentum are conserved in this scheme.  Staying in two dimensions in \S\ref{sec3}, we transform the equations of motion into a system of first order ODEs for the polar angles, angular momenta and eccentricity vectors $\vec{e}_i$ of the three sectors $i=1,2,3$ of the model.  We also demonstrate how the constraint can be employed to remove the third sector entirely, leaving equations for only $i=1,2$, corresponding to the planets.  Auxiliary vectors $\vec{f}_i=\hat{r}_i+\vec{e}_i$ for each sector are introduced in \S\ref{auxsec}, which make a considerable simplification to the algebraic form of the dynamical equations.  The geometrical properties of the auxiliary vector is explored.  We also present alternate forms of these equations in \S\ref{others}, in terms of the polar representations $(e_i,\beta_i)$, $(f_i,\psi_i)$ of eccentricity and auxiliary vectors, respectively.

In Part~\ref{part2} we return to three dimensions, and move toward the perturbational analysis of this problem.  In \S\ref{delaunay} we begin with the geometric definition and construction of the Delaunay action-angle variables for the two-body problem, with particular focus on the mean anomaly. In \S\ref{hamil} we present Hamilton's equations for the problem, in terms of the the perturbing function $\mathbf{R}=-\vec{\lambda}\cdot(\vec{r}_1+\vec{r}_2+\vec{r}_3)$, where $\vec{\lambda}$ is the Lagrange multiplier, $\vec{r}_1+\vec{r}_2+\vec{r}_3=0$ being the constraint.  Finally, in \S\ref{kam} we present the basic approach and the setup of Kolmogorov-Arnold-Moser (KAM) theory
for this problem.

\part{The Planetary Three Body Problem}\label{part1}

Let the masses of light bodies be $m_J, m_S$  respectively, and $M$ the mass of a heavier body.  Let $\mu=m_S/m_J$, expected to be $\mathcal{O}(1)$, while $\epsilon=m_J/M$ is small $\epsilon\ll 1$.  For Jupiter, Saturn and the sun these are
	\begin{equation}
		\begin{array}{cc}
			\begin{array}{l}
				M=1.989\times10^{30}\text{ kg} \\ m_J=1.898\times10^{27}\text{ kg} 	\\ m_S=5.683\times10^{26}\text{ kg}
			\end{array} &
			\begin{array}{l} \epsilon=9.54..\times10^{-4}\approx10^{-3} \\ \mu=0.2994..\approx0.3 \end{array}
		\end{array}
	\end{equation}
Let the positions of the bodies in an inertial reference
frame be $\vec{X}$ for mass $M$, and $\vec{x}_J,\vec{x}_S$ for $m_J,m_S$.  Then the Lagrangian is
	\begin{equation}
			\mathcal{L}_\text{in}=\frac{1}{2}\left(M\left\|\dot{\vec{X}}\right\|^2+m_J\norm{\dot{\vec{x}}_J}^2+m_S\norm{\dot{\vec{x}}_S}^2\right)+G\left(\frac{M m_J}{\|\vec{x}_J-\vec{X}\|}+\frac{M m_S}{\norm{\vec{x}_S-\vec{X}}}+\frac{m_J m_S}{\norm{\vec{x}_S-\vec{x}_J}}\right)
	\end{equation}
We will make a change of variables to: the center of mass and heliocentric coordinates
	\begin{align}
	 	\vec{R}&=\frac{M\vec{X}+m_J\vec{x}_J+m_S\vec{x}_S}{M_\Sigma} \\
	 	\vec{r}_J&=\vec{x}_J-\vec{X} \\
	 	\vec{r}_S&=\vec{X}-\vec{x}_S
	\end{align}
\noindent where $M_\Sigma=M+m_J+m_S$ is the total mass.  With these the Lagrangian becomes
	\begin{align}
		\mathcal{L}_\text{in}=&\frac{1}{2}M_\Sigma\norm{\dot{\vec{R}}}^2+\frac{1}{2}\frac{M}{M_\Sigma}\left\lbrace m_J\norm{\dot{\vec{r}}_J}^2+m_S\norm{\dot{\vec{r}}_S}^2+\frac{m_J m_S}{M}\norm{\dot{\vec{r}}_J+\dot{\vec{r}}_S}^2\right\rbrace \nonumber\\
		&+\frac{m_JM}{M_\Sigma}\cdot 	GM_\Sigma\left(r_J^{-1}+\frac{m_S/m_J}{r_S}+\frac{m_S/M}{\norm{\vec{r}_J+\vec{r}_S}}\right) \label{lag}
	\end{align}
We see the center-of-mass coordinate $\vec{R}$ is cyclic; it's dynamics decouple from the other variables, and so can be ignored.  We will consequently drop the first term of (\ref{lag}).  Then taking a factor $m_J$ out of the kinetic terms, we can see both remaining 
terms are proportional to $\tilde{m}_J=\frac{M}{M_\Sigma}m_J$, which we may regard as a reduced mass of `Jupiter'. Then a reduced and normalized Lagrangian is
	\begin{align}
		\mathbf{L}_\text{in}=\frac{\mathcal{L}_\text{in}}{\tilde{m}_J}=\frac{1}{2}\left(\norm{\dot{\vec{r}}_J}^2+\mu\norm{\dot{\vec{r}}_S}^2+\epsilon\mu\norm{\dot{\vec{r}}_J+\dot{\vec{r}}_S}^2\right)+\alpha\left(r_J^{-1}+\mu r_S^{-1}+\frac{\epsilon\mu}{\norm{\vec{r}_J+\vec{r}_S}}\right)
	\end{align}
\noindent where $\alpha=GM_\Sigma$ is the \textit{gravitational parameter}.  Often $\alpha$ is taken to be 1.  However, we will work in unit's of Jupiter's average distance and Jupiter's year, so that we should take $\alpha=R^3\omega^2=4\pi^2$.

At this stage we recognize 1) that the first and second terms of each parentheses, what can be called the `Jupiter' and `Saturn' terms, are just like the terms for the displacement vector of a two body problem, with large central mass at the origin, and 2) that the third terms, proportional to $\epsilon\mu$, also look like two-body problem terms, but with displacement vector $\vec{r}_{SJ}=\pm(\vec{r}_J+\vec{r}_S)$.  Taking the minus-sign option, which gives $\vec{r}_{SJ}=\vec{x}_S-\vec{x}_J$, we find we are dealing with a Lagrangian for three independent two-body problems, with vectors $\vec{r}_J,\vec{r}_S$ and $\vec{r}_{SJ}$ that satisfy the condition $\vec{r}_J+\vec{r}_S+\vec{r}_{SJ}=0$.  Looking back at our definition of these vectors in terms of inertial coordinates $\vec{X},\vec{x}_J,\vec{x}_S$, the constraint is satisfied identically:
	\begin{equation}
		(\vec{x}_J-\vec{X})+(\vec{X}-\vec{x}_S)+(\vec{x}_S-\vec{x}_J)\equiv0
	\end{equation}
\noindent We see that the three vectors are the sides of the triangle that the three bodies form.  The system is analogous to three bodies, of relative masses $1,\mu$ and $\epsilon\mu$ with respect to the first mass, all orbiting a central, stationary mass located at the origin, such that the gravitational parameter for each two-body problem is $\alpha=GM_\Sigma$.  Of particular note is that these three supposed bodies orbiting a central mass do not gravitate to each other.  The constraint is maintained by forces that are sourced by Lagrange multipliers.  We thus consider the modified Lagrangian
	\begin{equation}
		\mathbf{L}_\lambda=\sum_{i}\left\lbrace\mu_i\left(\frac{1}{2}\norm{\vec{r}_i}^2+\alpha/r_i\right)+\vec{\lambda}\bm{\cdot}\vec{r}_i\right\rbrace
	\end{equation}
\noindent where the sum is on $i=J,S,SJ$ or simply $i=1,2,3$, and $\mu_i=1,\mu,\epsilon\mu$.  The additional terms are $\vec{\lambda}\bm{\cdot}\sum_{i=1}^3\vec{r}_i$, so that the constraint equation is
	\begin{equation}
		\sum_{i=1}^3\vec{r}_i\,=0 \label{cond}
	\end{equation}
\noindent The Euler-Lagrange equations are
	\begin{equation}
		\mu_i\ddot{\vec{r}}_i=-\mu_i\frac{\alpha\,\vec{r}_i}{r_i^3}+\vec{\lambda}\,,\quad i=1,2,3. \label{ELeq}
	\end{equation}
\noindent The dynamical version of the constraint equation is 
	\begin{equation}
		\sum_{i=1}^3\ddot{\vec{r}}_i\,=0
	\end{equation}
\noindent If, in addition to this, we have initial conditions that satisfy both
	\begin{equation}
		\sum_{i=1}^3\vec{r}_i\,=0\text{ and }\sum_{i=1}^3\dot{\vec{r}}_i\,=0
	\end{equation}
\noindent then the condition (\ref{cond}) will be satisfied for all time.  This allows us to solve for the Lagrange multipliers by taking linear combination of the equations (\ref{ELeq}). We have
	\begin{equation}
		0=\sum_{i=1}^3\ddot{\vec{r}}_i=-\alpha\left(\sum_{i=1}^3\frac{\hat{r}_i}{r_i^2}\right)+\left(\sum_{i=1}^3\frac{1}{\mu_i}\right)\vec{\lambda}
	\end{equation}
\noindent and thus
	\begin{equation}
		\vec{\lambda}=\alpha\delta\sum_{i=1}^3\frac{\hat{r}_i}{r_i^2} \label{mult}
	\end{equation}
\noindent where the coefficient $\delta$ is the reciprocal of the sum of reciprocal masses
	\begin{equation}
		\delta=\left(\sum_{i=1}^3\frac{1}{\mu_i}\right)^{-1}=\frac{\epsilon\mu}{1+\epsilon+\epsilon\mu}=\frac{m_S}{M_\Sigma}=\frac{\tilde{m}_S}{M}=2.8536..\times 10^{-4}
	\end{equation}
The equations (\ref{ELeq}) are 
	\begin{equation}
		\ddot{\vec{r}}_i=A_{ij}\frac{\hat{r}_j}{r_j^2}\quad\text{(summation on $j$)}\label{eqs}
	\end{equation}
\noindent where the matrix $A$ of coefficients is
	\begin{align}
		A&=\alpha\left(\begin{array}{ccc}
			-\frac{1+\epsilon}{1+\epsilon+\epsilon\mu} & \delta & \delta \\
			\frac{\epsilon}{1+\epsilon+\epsilon\mu} & -\frac{1+\epsilon\mu}{1+\epsilon+\epsilon\mu} & \frac{\epsilon}{1+\epsilon+\epsilon\mu} \\
			(1+\epsilon+\epsilon\mu)^{-1} & (1+\epsilon+\epsilon\mu)^{-1} & -\frac{\epsilon(1+\mu)}{1+\epsilon+\epsilon\mu}
		\end{array}\right) \\
		&=G\left(\begin{array}{ccc}
			-(M+m_J) & m_S & m_S \\ m_J & -(M+m_S) & m_J \\ M & M & -(m_J+m_S)
		\end{array}\right).
	\end{align}
\noindent That each column sums to $0$ corresponds to $\ddot{\vec{r}}_1+\ddot{\vec{r}}_2+\ddot{\vec{r}}_3=0$.

\section{Constants of Motion}\label{constants}

\subsection{Energy}

The conjugate linear momenta are
\begin{equation}
		\vec{p}_i=\nabla_{\dot{\vec{r}}_i}\mathbf{L}_\lambda=\mu_i\dot{\vec{r}}_i
	\end{equation}
and by Legendre transform, the Hamiltonian is
	\begin{align}
		\mathbf{H}_\lambda&=\sum_i\vec{p}_i\bm{\cdot}\dot{\vec{r}}_i\,-\mathbf{L}_\lambda \\
		 &=\sum_i\left\lbrace \frac{p_i^2}{\mu_i}-\left(\frac{1}{2}\frac{p_i^2}{\mu_i}+\mu_i\frac{\alpha}{r_i}+\vec{\lambda}\bm{\cdot}\vec{r}_i\right)\right\rbrace \\
	 	&=\sum_i\left\lbrace \frac{p_i^2}{2\mu_i}-\mu_i\frac{\alpha}{r_i}-\vec{\lambda}\bm{\cdot}\vec{r}_i \right\rbrace
	\end{align}
\noindent thus when the constraint (\ref{cond}) is satisfied, the following (reduced) energy is conserved
	\begin{equation}
		\xi=\frac{E}{\tilde{m}_J}=\sum_i\mu_i\left(\frac{1}{2}\norm{\dot{\vec{r}}_i}^2-\alpha/r_i\right)=\sum_i \mu_i \xi_i
	\end{equation}
\noindent where $\xi_i=\norm{\dot{\vec{r}}_i}^2/2-\alpha/r_i$ is the specific energy for each sector of the model.

\subsection{Angular Momentum}

Now we put the dynamical vectors $\vec{r}_i$ in polar coordinates.  Up to this point, we could have been in three dimensions, but for now we work in two dimensions.  Each vector is $\vec{r}_i(t)=r_i(t)\hat{\theta}_i(t)$, where $\hat{r}(\theta)=(\cos\theta,\,\sin\theta)^\text{T}$
	\begin{align}
		&\norm{\dot{\vec{r}}_i}^2=\dot{r}_i^2+r_i^2\dot{\theta}_i^2 \\
		&\implies\quad\text{momenta conjugate to }\theta_i\text{ are } \nonumber\\
		&h_i=\frac{\partial\,\mathbf{L}_\lambda}{\partial\dot{\theta}_i}=\mu_i r_i^2\dot{\theta}_i \\
		&\text{and the Euler-Lagrange equations are } \nonumber\\
		&\dot{h}_i=\frac{\partial\,\mathbf{L}_\lambda}{\partial\theta_i}=\frac{\partial}{\partial\theta_i}\left(\vec{\lambda}\bm{\cdot}\vec{r}_i\right) \\
		&\hphantom{\dot{h}_i}=\vec{\lambda}\bm{\cdot}r_i\hat{\theta}_i \\
		&\hphantom{\dot{h}_i}=\vec{\lambda}\bm{\cdot}r_iT\hat{r}_i \\
		&\hphantom{\dot{h}_i}=\vec{\lambda}\bm{\cdot}T\vec{r}_i
	\end{align}
\noindent where $\hat{\theta}$ is the vector function $\hat{\theta}(\varphi)=(-\sin\varphi,\,\cos\varphi)^\text{T}$, $\hat{\theta}_i$ is the evaluation $\hat{\theta}(\theta_i(t))$, and $T$ is the $2\times2$ matrix $T=\left[\begin{array}{lr} 0 & -1 \\ 1 & 0 \end{array}\right]$, which is ccw-rotation by $\pi/2$, so that $\hat{\theta}=T\hat{r}$.  This shows that the total angular momentum satisfies
	\begin{equation}
		\dot{h}=\sum_{i=1}^3\dot{h}_i=\vec{\lambda}\bm{\cdot}T\sum_i\vec{r}_i=0\text{ by the constraint}.
	\end{equation}
\noindent The presence of the rotation matrix $T$ in $\dot{h}$ shows that conservation of angular momentum is due to the fact that the system as a whole is invariant under global rotation.

Going forward, it will be best to use specific angular momenta
	\begin{equation}
		\ell_i=h_i/\mu_i=r_i^2\dot{\theta}_i
	\end{equation}
\noindent which gives
	\begin{equation}
		\dot{\ell}_i=\frac{1}{\mu_i}\vec{\lambda}\bm{\cdot}r_i\hat{\theta}
	\end{equation}
Using the Lagrange multiplier (\ref{mult}), this is
	\begin{align}
		\dot{\ell}_i&=\frac{\alpha\delta}{\mu_i}\left(\sum_j\frac{\hat{r}_j}{r_j^2}\right)\bm{\cdot}r_i\hat{\theta}_i \\
		\therefore \dot{\ell}_i&=\frac{\alpha\delta}{\mu_i}r_i\sum_{j\neq i}r_j^{-2}\sin(\theta_j-\theta_i) \label{torque}
	\end{align}
\noindent we shall label these functions $\dot{\ell}_i=\tau_i=\tau_i(r_j,\theta_j,\ell_j)$; $\tau$ for torque.  Notice that these equations are perturbations for $i=1,2$, since $\mu_{1,2}\sim\mathcal{O}(1)$, but this is not the case for $i=3$; $\mu_3=\epsilon\mu\sim\mathcal{O}(\delta)$, so the coefficient in the equation for $\dot{\ell}_3$ is $\mathcal{O}(1)$.

\subsubsection{Angular Momentum in Three Dimensions}

In three dimensions, we have the vector (total) angular momentum
	\begin{equation}
		\vec{h}=\sum_{i}\vec{r}_i\times\vec{p}_i=\sum_i\mu_i\vec{r}_i\times\dot{\vec{r}}_i
	\end{equation}
\noindent the time-derivative of which is
	\begin{align}
		\dot{\vec{h}}&=\sum_i\vec{r}_i\times\dot{\vec{p}}_i \\
		&=\sum_i \vec{r}_i\times\left(-\mu_i\alpha\frac{\hat{r}_i}{r_i^2}+\vec{\lambda}\right) \\
		&=\left(\sum_i\vec{r}_i\right)\times\vec{\lambda}
	\end{align}
\noindent so angular momentum is conserved by the constraint.

\section{Returning to the Equations of Motion}\label{sec3}

	\begin{equation}
		\ddot{\vec{r}}_i=A_{ij}\frac{\hat{r}_j}{r_j^2} \label{sys}
	\end{equation}
With $\vec{r}_i(t)=r_i(t)\hat{\theta_i(t)}$, we develop these equations in polar coordinates.  We will work out the $\hat{r}_i$- and $\hat{\theta}_i$-components for the $i^\text{th}$ equation of (\ref{sys2}).
	\begin{align}
		&\dot{\vec{r}}_i=\dot{r}_i\hat{r}_i+r_i\dot{\theta}_i\hat{\theta}_i \\
		&\text{so }\norm{\dot{\vec{r}}_i}^2=\dot{r}_i^2+r_i^2\dot{\theta}_i^2\,\,,\,\text{ and } \nonumber\\
		&\ddot{\vec{r}}_i=(\ddot{r}_i-r_i\dot{\theta}_i^2)\hat{r}_i+(r_i\ddot{\theta}_i+2\dot{r}_i\dot{\theta}_i)\hat{\theta}_i\,.
	\end{align}
\noindent The $\hat{\theta}_i$-component of $\ddot{\vec{r}}_i$ is nothing other than $\dot{\ell}_i/r_i$; indeed the $\hat{\theta}_i$-components of (\ref{sys}) are just the torque equations (\ref{torque}) derived above.  That leaves us with the $\hat{r}_i$-components
	\begin{equation}
		\ddot{r}_i-r_i\dot{\theta}^2_i=\left(\sum_{j=1}^3A_{ij}\frac{\hat{r}_j}{r_j^2}\right)\bm{\cdot}\hat{r}_i
	\end{equation}
\noindent with $\dot{\theta}_i=\ell_i/r_i^2$ and $A_{ii}=\alpha(-1+\delta/\mu_i)$, while $A_{ij}=\alpha\delta/\mu_i$ for $j\neq i$, this is
	\begin{align}
		&\ddot{r}_i=\frac{\ell_i^2}{r_i^3}+\frac{A_{ii}}{r_i^2}+\frac{\alpha\delta}{\mu_i}\sum_{j\neq i}\frac{\cos(\theta_j-\theta_i)}{r_j^2} \\
		\therefore\quad&\ddot{r}_i=\frac{\ell_i^2}{r_i^3}-\alpha r_i^{-2}+\frac{\alpha\delta}{\mu_i}\sum_{j=1}^3\frac{\cos(\theta_j-\theta_i)}{r_j^2}
	\end{align}
Here, in analogy to the definition we have for $\tau_i=\dot{\ell}_i=(\alpha\delta/\mu_i)\,r_i\sum_j\sin(\theta_j-\theta_i)/r_j^2$, we define three functions $\sigma_i$ as
	\begin{align}
		\sigma_i&=\frac{1}{\mu_i}\vec{\lambda}\bm{\cdot}\vec{r}_i \nonumber \\
		&=\frac{\alpha\delta}{\mu_i}\,r_i\sum_{j=1}^3\frac{\cos(\theta_j-\theta_i)}{r_j^2}\label{sigmas}
	\end{align}
\noindent then what we have is 
	\begin{equation}
		\ddot{r}_i=\frac{\ell_i^2}{r_i^3}-\alpha r_i^{-2}+\frac{\sigma_i}{r_i} \label{accel}
	\end{equation}
\noindent and define the right-hand sides as functions $a_i=a_i(\theta,\ell,r)$

\subsection{Eccentricity or Laplace-Runge-Lenz Vectors}

At this stage, we may formulate our differential equations as a system of 12 first order ODEs
	\begin{align}
		\dot{\theta_i}&=\Omega_i=\ell_i/r_i^2 \label{sys01}\\
		\dot{\ell}_i&=\tau_i=\frac{\alpha\delta}{\mu_i}r_i\sum_{j\neq i}r_j^{-2}\sin(\theta_j-\theta_i) \\
		\dot{r}_i&=v_i \\
		\dot{v}_i&=a_i=\frac{\ell_i^2}{r_i^3}-\alpha r_i^{-2}+\frac{\sigma_i}{r_i} \label{sys04}
	\end{align}
Now we introduce a change of variables from $(r_i,v_i)$ to \textit{eccentricity vectors}, in each sector
	\begin{equation}
		\vec{e}=\dot{\vec{r}}\times\vec{\ell}/\alpha-\hat{r} \label{evec}.
	\end{equation}
\noindent This is a normalization of the Laplace-Runge-Lenz vector
	\begin{equation}
		\vec{A}=\vec{p}\times\vec{L}-m^2\alpha\,\hat{r}=m^2\alpha\,\vec{e}
	\end{equation}
\noindent where $\vec{p}$ is linear momentum and $\vec{L}$ is the (dimensionfull) angular momentum.  In the two-body problem, these vectors are constants of motion.  The eccentricity vector has magnitude equal to that of the eccentricity of the Keplerian orbit and points in the direction of periapsis.  Working in two dimensions, our specific angular momenta are out of the plane, $\vec{\ell}=\ell\hat{z}$, so that in polar coordinates
	\begin{align}
		\vec{e}&=(v\,\hat{r}+r\,\dot{\theta}\,\hat{\theta})\times(\ell\,\hat{z}/\alpha)-\hat{r} \\
		&=\left(\frac{\ell^2/\alpha}{r}-1\right)\hat{r}-\ell v/\alpha\,\hat{\theta}
	\end{align}
\noindent (where $\ell/r^2$ has been substituted for $\dot{\theta}$).  This defines the eccentricity vector in terms of it's polar components
	\begin{equation}
		e^r=\frac{\ell^2/\alpha}{r}-1,\quad e^\theta=-\ell\,v/\alpha
	\end{equation}
\noindent the reverse change of variables being
	\begin{align}
		r&=\frac{\ell^2/\alpha}{1+e^r} \label{osculating} \\
		v&=-\alpha\frac{e^\theta}{\ell} \label{etheta}.
	\end{align}
\noindent Equation (\ref{osculating}) is precisely the form of a Keplerian elliptic orbit if $\ell$ and $\vec{e}$ are constant, as in the two-body problem, since
	\begin{align}
		r=\frac{\ell^2/\alpha}{1+e^r}&=\frac{\ell^2/\alpha}{1+\vec{e}\bm{\cdot}\hat{r}} \\
		&=\frac{\ell^2/\alpha}{1+e^x\cos\theta+e^y\sin\theta} \\
		&=\frac{\ell^2/\alpha}{1+e\cos(\theta-\beta)}
	\end{align}  
\noindent where $\beta$ is the angle of $\vec{e}\,$ from the positive $x$-axis, called the \textit{longitude of periapsis}.  Equation $(\ref{osculating})$ thus describes the \textit{osculating orbit} to the trajectory, which is the elliptic orbit that a body would follow given it's instantaneous angular momentum and eccentricity.  As a further note, if we take the time-derivative of (\ref{osculating}), writing $\tau$ for $\dot{\ell}$, we find
	\begin{equation}
		v=\dot{r}=\frac{1}{\alpha}\left(\frac{2\ell\tau}{1+e^r}-\frac{\ell^2}{(1+e^r)^2}\left(\dot{\vec{e}}\bm{\cdot}\hat{r}+\frac{\ell}{r^2}\vec{e}\bm{\cdot}\hat{\theta}\right)\right)=\frac{1}{\alpha}\left(\frac{2\ell\tau}{1+e^r}-\frac{r^2}{\ell^2}\dot{\vec{e}}\bm{\cdot}\hat{r}\right)-\alpha\frac{e^\theta}{\ell} \label{diff}
	\end{equation} 
\noindent which reduces to precisely (\ref{etheta}) if $\tau$ and $\dot{\vec{e}}$ are 0.

\subsection{System of Equations using Eccentricity vectors}

We can differentiate the definitions (\ref{evec}) and use the system (\ref{sys01}-\ref{sys04}) to derive first order differential equations for the eccentricity vectors $\dot{\vec{e}}_i=\dots(\theta,\ell,\vec{e}\,)$.  First working in one sector (no indices $i$), writing $\tau$ for $\dot{\ell}$, we have
	\begin{align}
		\dot{\vec{e}}&=\frac{1}{\alpha}\left(\frac{2\ell\tau}{r}-\cancel{\frac{\ell^2v}{r}}\,\right)\hat{r}+\left(\frac{\ell^2/\alpha}{r}-1\right)\frac{\ell}{r^2}\hat{\theta}-(\tau\,v+\ell\,\ddot{r})\,\hat{\theta}/\alpha+\cancel{\frac{\ell v}{\alpha}\frac{\ell}{r^2}\hat{r}} \\
		&= \frac{2\ell\tau}{\alpha\,r}\hat{ 	r}-\frac{\ell}{\alpha}\left(\ddot{r}-\frac{\ell^2}{r^3}+\alpha r^{-2}+v\frac{\tau}{\ell}\right)\hat{\theta}. \\
	\intertext{Substituting $v=-\alpha e^\theta/\ell$ and $\ddot{r}=a=\ell^2/r^3-\alpha r^{-2}+\sigma/r$ gives}
		\dot{\vec{e}}&=\frac{2\ell\tau}{\alpha\,r}\hat{r}+\left(\frac{\tau}{\ell}\,e^\theta-\frac{\ell\sigma}{\alpha\,r}\right)\hat{\theta}. \\
	\intertext{Pair half of the first term with the second $\hat{\theta}$-term}
		\dot{\vec{e}}&=\frac{\ell\tau}{\alpha\,r}\hat{r}+\frac{\tau}{\ell}e^\theta\hat{\theta}+\frac{\ell}{\alpha\,r}\left(\tau\hat{r}-\sigma\hat{\theta}\right). \\
	\intertext{In the first term we substitute $\ell/\alpha\,r=(1+e^r)/\ell$}
		\dot{\vec{e}}&=\frac{\tau}{\ell}(1+e^r)\hat{r} + \frac{\tau}{\ell}e^\theta\hat{\theta}+\frac{\ell}{\alpha\,r}\left(\tau\hat{r}-\sigma\hat{\theta}\right) \\ 
		&=\frac{\tau}{\ell}\left(\hat{r}+e^r\hat{r}+e^\theta\hat{\theta}\right)+\frac{\ell}{\alpha\,r}\left(\tau\hat{r}-\sigma\hat{\theta}\right). \\
	\intertext{Thus we find}
		\dot{\vec{e}}&=\frac{\tau}{\ell}\left(\vec{e}+\hat{r}\right)+\frac{\ell}{\alpha\,r}\left(\tau\,\hat{r}-\sigma\,\hat{\theta}\right).
	\end{align}
\noindent The combination $(\tau\,\hat{r}-\sigma\,\hat{\theta})/r$ is, in each sector
	\begin{align}
		(\tau_i\,\hat{r}_i-\sigma_i\,\hat{\theta}_i)/r_i=&\left(\frac{\alpha\delta}{\mu_i}\sum_{j=1}^3\frac{\sin(\theta_j-\theta_i)}{r_j^2}\right)\left[\begin{array}{c} \cos\theta_i \\ \sin\theta_i \end{array}\right] \nonumber \\
		&-\left(\frac{\alpha\delta}{\mu_i}r_i\sum_{j=1}^3\frac{\cos(\theta_j-\theta_i)}{r_j^2}\right)\left[\begin{array}{c} -\sin\theta_i \\ \cos\theta_i \end{array}\right] \\
		=&\frac{\alpha\delta}{\mu_i}\,\sum_{j=1}^3r_j^{-2}\left[\begin{array}{c} \sin\theta_j \\ -\cos\theta_j \end{array}\right] \nonumber \\
		=&-\frac{\alpha\delta}{\mu_i}\,\sum_{j=1}^3r_j^{-2}\hat{\theta}_j=-\frac{\alpha\delta}{\mu_i}\,\sum_{j=1}^3r_j^{-2}\,T\hat{r}_j  \\
		\therefore\,(\tau_i\,\hat{r}_i-\sigma_i\,\hat{\theta}_i)/r_i=&-T\vec{\lambda}/\mu_i \label{Tlam}
	\end{align}
So finally, our differential equations for $\dot{\vec{e}}_i$ are
	\begin{equation}
		\dot{\vec{e}}_i=\frac{\tau_i}{\ell_i}\left(\vec{e}_i+\hat{r}_i\right)-\frac{\ell_i}{\alpha\mu_i}\,T\vec{\lambda}. \label{ecc.vel}
	\end{equation}
\noindent Observe that these equations are $\propto\,\delta/\mu_i$, so that $\dot{\vec{e}}_i$ are $\mathcal{O}(\delta)$ for $i=1,2$, i.e. Jupiter and Saturn, while $\dot{\vec{e}}_3$ is $\mathcal{O}(1)$.  The system of twelve first order ODEs for the variables $(\theta_i,\ell_i,\vec{e}_i)$ is
	\begin{align}
		\dot{\theta_i}&=\Omega_i=\ell_i/r_i^2 \label{sys1}\\
		\dot{\ell}_i&=\tau_i=\frac{\alpha\delta}{\mu_i}r_i\sum_{j\neq i}r_j^{-2}\sin(\theta_j-\theta_i) \label{sys2}\\
		\dot{\vec{e}}_i&=\frac{\tau_i}{\ell_i}\left(\vec{e}_i+\hat{r}_i\right)-\frac{\ell_i}{\alpha\mu_i}T\vec{\lambda}\quad\text{for }i=1,2,3 \label{sys3} \\
		&\text{where }r_i=\ell_i^2/\left[\alpha(1+e^r_i)\right] \nonumber
	\end{align}
\noindent Mixing between the sectors enters through the torques $\tau_i$ and $\vec{\lambda}$.





\subsection{Reduction by the Constraint to Two Sectors}

Given that solutions will satisfy the constraint $0=\sum_i\vec{r}_i$, we can write down algebraic/trigonometric expressions for the variables of the 3rd sector $\vec{r}_3=-\vec{r}_1-\vec{r}_2$ in terms of the 1st and 2nd sectors.  Substituting these relations into the $i=1,2$ equations
would leave the system (\ref{sys1}-\ref{sys3}) for $i=1,2$ only, and the equations for $\dot{\ell}_i,\dot{\vec{e}}_i$ would all be $\mathcal{O}(\delta)$.

In particular, if $\vec{r}_3=-\vec{r}_1-\vec{r}_2$, then we can write expressions for the radius and trig ratios of the argument of $\vec{r}_3$ in terms of those of $\vec{r}_1,\vec{r}_2$.  First of all, by the cosine-law
	\begin{equation}
		r_3^2=r_1^2+r_2^2+2 r_1r_2\cos(\theta_2-\theta_1)\label{r3}
	\end{equation}
\noindent and then from basic trigonometry, we can express $\cos\theta_3,\sin\theta_3$ as follows
	\begin{align}
		&\cos\theta_3=-\frac{r_1\cos\theta_1+r_2\cos\theta_2}{r_3}\label{c3} \\
		&\sin\theta_3=-\frac{r_1\sin\theta_1+r_2\sin\theta_2}{r_3}\label{s3}
	\end{align}
These may then be worked into the equations (\ref{sys2},\ref{sys3}), still using $r_i=\alpha^{-1}\ell_i^2/(1+e^r_i)$ but now only for $i=1,2$.  This results in a system of first order ODEs $\dot{\theta}_1=\Omega_1,\,\dot{\theta}_2=\Omega_2,\,\dot{\ell}_1=\tau_1,\,\dot{\ell}_2=\tau_2,\dot{\vec{e}}_1=\vec{v}_1,\dot{\vec{e}}_2=\vec{v}_2$ where the angular velocities $\Omega_i$ are $\mathcal{O}(1)$, but the torques $\tau_i$ and eccentricity-velocities $\vec{v}_i$ are $\mathcal{O}(\delta)$, facilitating a multiple-scales analysis.

\section{Auxiliary vectors $\vec{f}=\hat{r}+\vec{e}$}\label{auxsec}

The expressions for the right-hand-sides of (\ref{sys1}-\ref{sys3}), especially the torques and eccentricity-velocities, in terms of the variables $(\theta,\ell,e^r,e^\theta)$, are very large and cumbersome.  Written as rational functions, the numerators and denominators expanded out, are respectively 16 and 9 terms for the torques and 222 and 9 terms for the eccentricity-velocities, not including terms within $3/2$-roots of the denominators.

The prevalence of the combination $\hat{r}_i+\vec{e}_i$ presents an opportunity for simplification.  Observe that the denominator of the expression for the osculating orbit (\ref{osculating}) is the $\hat{r}_i$-component of this vector
	\begin{equation}
		r=\frac{\ell^2/\alpha}{1+e^r}=\frac{\ell^2/\alpha}{(\hat{r}+\vec{e}\,)\cdot\hat{r}}.
	\end{equation}
\noindent The combination $\hat{r}_i+\vec{e}_i$ also occurs in the DEs for eccentricity vectors (\ref{ecc.vel}). If we change variables from eccentricity vectors $\vec{e}_i$ to these auxiliary vectors $\vec{f}_i=\hat{r}_i+\vec{e}_i$, the modified equations become
	\begin{align}
		r_i&=\frac{\ell_i^2}{\alpha\,f_i^r}=\frac{\ell_i^2}{\alpha\,\vec{f}_i\cdot\hat{r}_i}=\frac{\ell_i^2}{\alpha\,f_i\cos(\psi_i-\theta_i)} \\
		\dot{\vec{f}}_i&=\frac{\tau_i}{\ell_i}\vec{f}_i-\frac{\ell_i}{\alpha\mu_i}(T\vec{\lambda})+\frac{d}{dt}\hat{r}_i \\
		&=\frac{\tau_i}{\ell_i}\vec{f}_i-\frac{\ell_i}{\alpha\mu_i}(T\vec{\lambda})+\Omega_i\,\hat{\theta}_i \label{f_de}
	\end{align}
\noindent where $f_i=\norm{\vec{f}_i}$ and $\psi_i$ is the argument of $\vec{f}_i$ (such that $\vec{f}_i/f_i=\hat{r}(\psi_i)$).  Let the point be laboured, that given this definition of $\vec{f}_i$, the differential equation is
	\begin{equation}
		\dot{\vec{f}}_i=\frac{d}{dt}\hat{r}_i+\mathcal{O}(\delta).
	\end{equation}
\noindent So to leading order, the vector $\vec{f}_i$ will be very nearly equal to the unit vector $\hat{r}_i(t)=\hat{r}(\theta_i(t))$.  Indeed, the solution is exactly $\vec{f}_i=\hat{r}_i+\vec{e}_i$.  So solutions will have $f_i\sim 1$ and $\psi_i\sim\theta_i$ for small initial eccentricity, at least for a finite duration after initial conditions.

\subsection{Geometric relationship of the Auxiliary vectors}

	\begin{figure}[h!]
		\centering
		\subfigure[\label{aux1}]{\includegraphics[scale=0.5]{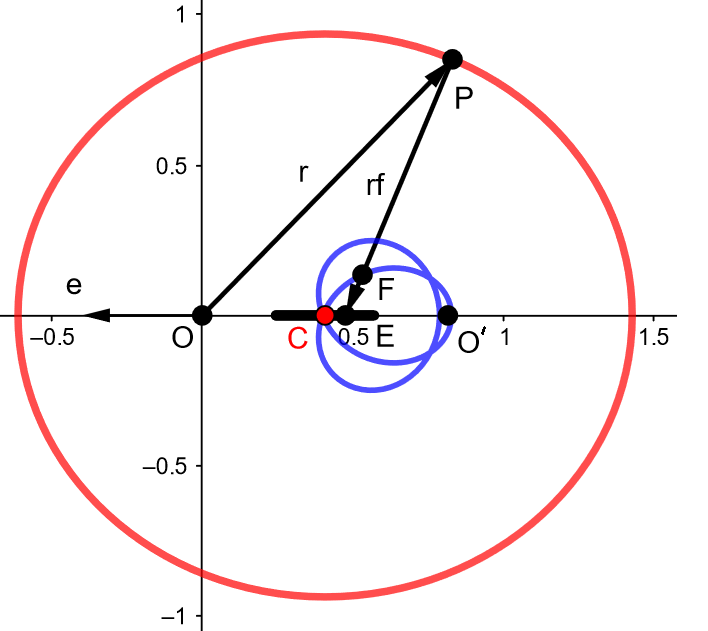}}
		\subfigure[\label{aux2}]{\includegraphics[scale=0.55]{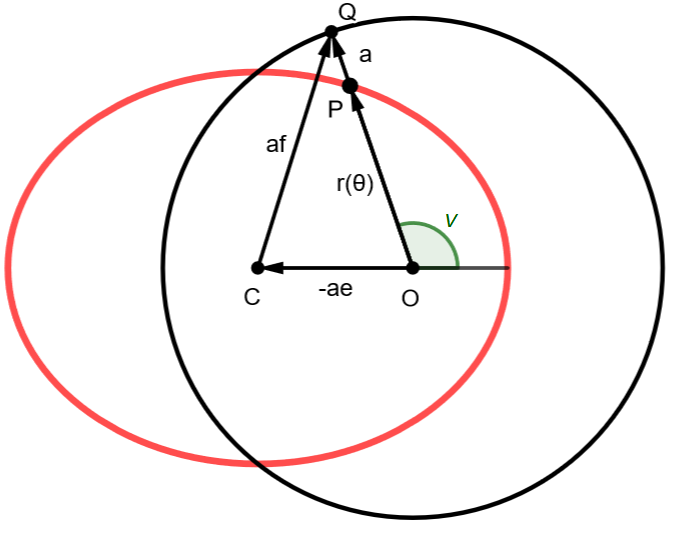}}
		\caption{These figures shows the elliptic geometry of the auxiliary vector. In a) $\vec{f}=\hat{r}+\vec{e}$, where $\vec{e}$ is the eccentricity vector, shown to the left of $A$.  This ellipse has an eccentricity of 0.44 and semi-major axis equal to 1.02.  The points $O,O^\prime$ are the foci, and $C$ is the center.  The point $P$ represents the location of a planet, so the vector from $O$ to $P$ is $\vec{r}$.  The point $E$ on the semi-major axis is the location of $\vec{r}-r\vec{f}=-r\vec{e}$.  The point $F$ on the blue curve is the coordinate of $\vec{r}-\vec{f}$, and the blue curve is the locus of such points, for all possible points $P$ on the ellipse.  The black interval along the $x$-axis is the locus of points $E$.  As a subset of the $x$-axis, it is the interval $[ae(1-e),ae(1+e)]$.  In b) the ellipse has eccentricity $e=0.6$ and $a=1$; the equation $a\hat{r}=-a\vec{e}+a\vec{f}$ is demonstrated, where $a\hat{r}=\vec{OQ}$ is the vector position of the orbit $\vec{OP}$ continued to circle about $O$, and $\vec{CQ}$ is the auxiliary vector $a\vec{f}$.}
	\end{figure}
For ellipses with even moderate eccentricity, up to $\sim0.5$, the vector $-\vec{f}=-\hat{r}-\vec{e}$ points, to leading order, from the position of the planet towards the center of the ellipse.  Indeed, the vector $r\vec{f}=\vec{r}+r\vec{e}$ is such that
	\begin{equation}
		\vec{r}-r\vec{f}=-r\vec{e},
	\end{equation}
\noindent while the coordinate of the center of the ellipse is $-a\vec{e}$, where $a$ is the semi-major axis, as demonstrated in Fig.~\ref{aux1}.  So the degree to which these coincide is the degree to which $r$ and $a$ agree.  In terms of eccentricity, semi-major axis and \textit{true anomaly} $\nu=\theta-\beta$, the relationship is
	\begin{equation}
		r=\frac{a(1-e^2)}{1+e\cos\nu}.
	\end{equation}
At minimum $re=ae(1-e)$, while the maximum value is $ae(1+e)$.  Thus the location $\vec{r}-r\vec{f}$ lies within a segment of the major-axis which is a length $a\,e^2$ on either side of the ellipse center, as shown in Fig.~\ref{aux1}. The error of $-\vec{f}$ pointing from the location of the planet to the center of the orbit is $\mathcal{O}(e^2)$.   Specifically, the difference in angle of $-r\vec{e}$ vs. $-a\vec{e}$ as seen from the position $\vec{r}$---in other words, the angle $\angle CPE$---is $e^2\sin\nu\cos\nu$ to leading order in $e$.

Moreover, if we consider the equation
	\begin{equation}
		a\hat{r}=-a\vec{e}+a\vec{f}, \label{interp}
	\end{equation}
\noindent then we see the following geometry: construct an circle around the focus of the ellipse, with radius $a$.  If we continue the vector $\vec{r}$ from the focus out to radius $a$, we reach the point $a\hat{r}$ on the circle.  The position of the ellipse centre from the focus is the first term $-a\vec{e}$.  Thus we see from (\ref{interp}) that the vector $a\vec{f}$ points from the centre of the ellipse to the position of the orbit projected radially to radius $a$, as shown in Fig.~\ref{aux2}.  

\subsection{The Dynamical System in terms of the Auxiliary vectors}

What might simplify the equations the most is to write the equations for $\dot{\vec{f}}_i$ in polar form, for the components $\dot{f}_i^r=\frac{d}{dt}(\hat{r}_i\cdot\vec{f}_i)$ and $\dot{f}_i^\theta=\frac{d}{dt}(\hat{\theta}_i\cdot\dot{\vec{f}}_i)$.  That is, we have
	\begin{align}
		\vec{f}_i&=f_i^r\hat{r}_i+f_i^\theta\hat{\theta}_i \\
		\text{and }\dot{\vec{f}}_i&=\dot{f}_i^r\hat{r}_i+f_i^r\left(\Omega_i\hat{\theta}_i\right) +\dot{f}_i^\theta\hat{\theta}_i+f_i^\theta\left(-\Omega_i\hat{r}_i\right) \\
		&=\left(\dot{f}_i^r-\Omega_i f_i^\theta\right)\hat{r}_i+\left(\dot{f}_i^\theta+\Omega_i f_i^r\right)\hat{\theta}_i
	\end{align}

The equation (\ref{f_de}) for $\dot{\vec{f}}_i$ becomes
	\begin{align}
		\dot{f}_i^r&=\frac{\tau_i}{\ell_i}f_i^r+\Omega_i 	f_i^\theta-\frac{\ell_i}{\alpha\mu_i}\left(\hat{r}_i\cdot T\vec{\lambda}\right) \\
		\dot{f}_i^\theta&=\frac{\tau_i}{\ell_i}f_i^\theta-\Omega_i f_i^r-\frac{\ell_i}{\alpha\mu_i}\left(\hat{\theta}_i\cdot T\vec{\lambda}\right)+\Omega_i.
	\end{align}
\noindent We know the components of $T\vec{\lambda}$ from (\ref{Tlam})\footnote{Multiplying (\ref{Tlam}) by $-T$, we can also find $\vec{\lambda}=\frac{\mu_i}{r_i}\left(\sigma_i\hat{r}_i+\tau_i\hat{\theta}_i\right)$.}
	\begin{align}
		\hat{r}_i\cdot(-T\vec{\lambda})=\mu_i\,\tau_i/r_i \\
		\hat{\theta}_i\cdot(-T\vec{\lambda})=-\mu_i\,\sigma_i/r_i
	\end{align}
\noindent which gives
	\begin{align}
		\dot{f}_i^r&=\frac{\tau_i}{\ell_i}f_i^r+\frac{\ell_i\tau_i}{\alpha\,r_i}+\Omega_i f_i^\theta \label{fcomp1} \\
		\dot{f}_i^\theta&=\frac{\tau_i}{\ell_i}f_i^\theta-\frac{\ell_i\sigma_i}{\alpha\,r_i}+\Omega_i(1-f_i^r) \label{fcomp2}.
	\end{align}

\subsection{Final substitutions}

Things may be made yet more concise.  The osculating orbits are given by $r_i=\ell_i^2/\alpha f_i^r$, and the prevalent coefficients in (\ref{fcomp1},\ref{fcomp2}) become 
	\begin{align}
		\Omega_i&=\frac{\ell_i}{r_i^2}=\alpha^2\frac{{f_i^r}^2}{\ell_i^3}\label{orbv} \\
		\frac{\ell_i}{\alpha\,r_i}&=\frac{f_i^r}{\ell_i} \label{foo}
	\end{align}
\noindent so that we find
	\begin{align}
		\dot{f}_i^r&=2\frac{\tau_i}{\ell_i}f_i^r+\alpha^2\frac{{f_i^r}^2}{\ell_i^3}f_i^\theta \label{fnew1} \\
		\dot{f}_i^\theta&=\left(-\frac{\sigma_i}{\ell_i}+\alpha^2\frac{{f_i^r}(1-f_i^r)}{\ell_i^3}\right)f_i^r+\frac{\tau_i}{\ell_i}f_i^\theta. \label{fnew2}
	\end{align}
For completeness, we give the expressions for $\sigma$ and $\tau$ in these terms
	\begin{align}
		\tau_i&=\frac{\alpha^2\delta}{\mu_i}\frac{\ell_i^2}{f_i^r}\sum_{j\neq i}\frac{{f_j^r}^2}{\ell_j^4}\sin(\theta_j-\theta_i) \label{taus} \\
		\sigma_i&=\frac{\alpha^2\delta}{\mu_i}\left[\frac{f_i^r}{\ell_i^2}+\frac{\ell_i^2}{f_i^r}\sum_{j\neq i}\frac{{f_j^r}^2}{\ell_j^4}\cos(\theta_j-\theta_i)\right].\label{sigs}
	\end{align}
\noindent Equations (\ref{taus},\ref{sigs}) are best read in the 3-sector version of this problem (that is, if one does not eliminate $r_3$ and $\theta_3$).  If one wishes to work in only two sectors $i=1,2$, the substitutions (\ref{r3},\ref{c3},\ref{s3}) should be made into (\ref{torque}) and (\ref{sigmas}).

We make the substitutions (\ref{orbv},\ref{foo}) into (\ref{ecc.vel}) and (\ref{f_de}) as well, which become
	\begin{align}
		\dot{\vec{e}}_i&=\frac{\tau_i}{\ell_i}(\vec{e}_i+\hat{r}_i)+\frac{f_i^r}{\ell_i}\left(\tau_i\hat{r}_i-\sigma_i\hat{\theta}_i\right) \label{ecc.vel_sub} \\
		\dot{\vec{f}}_i&=\dot{\vec{e}}_i+\Omega_i\hat{\theta}_i \nonumber \\
		&=\frac{\tau_i}{\ell_i}\vec{f}_i+\frac{f_i^r}{\ell_i}\left(\tau_i\hat{r}_i-\sigma_i\hat{\theta}_i\right)+\alpha^2\frac{{f_i^r}^2}{\ell_i^3}\hat{\theta}_i\label{f_de_sub}
	\end{align}

In full then, the system of equations, in terms of the polar components of the auxiliary vectors $f_i^r,f_i^\theta$ are
	\begin{align}
		\dot{\theta_i}&=\Omega_i=\frac{\alpha^2{f_i^r}^2}{l_i^3} \label{fs1} \\
		\dot{\ell}_i&=\tau_i=(\ref{taus}) \\
		\dot{f}_i^r&=2\frac{\tau_i}{\ell_i}f_i^r+\alpha^2\frac{{f_i^r}^2}{\ell_i^3}f_i^\theta  \\
		\dot{f}_i^\theta&=\left(-\frac{\sigma_i}{\ell_i}+\alpha^2\frac{{f_i^r}(1-f_i^r)}{\ell_i^3}\right)f_i^r+\frac{\tau_i}{\ell_i}f_i^\theta. \label{fs4}
	\end{align}

\noindent Numerical solutions of (\ref{fs1}-\ref{fs4}) for Jupiter's eccentricity vector are shown in Fig.

\begin{figure}[h!]
	\centering
	\subfigure[\label{ecc1}]{\includegraphics[scale=0.6]{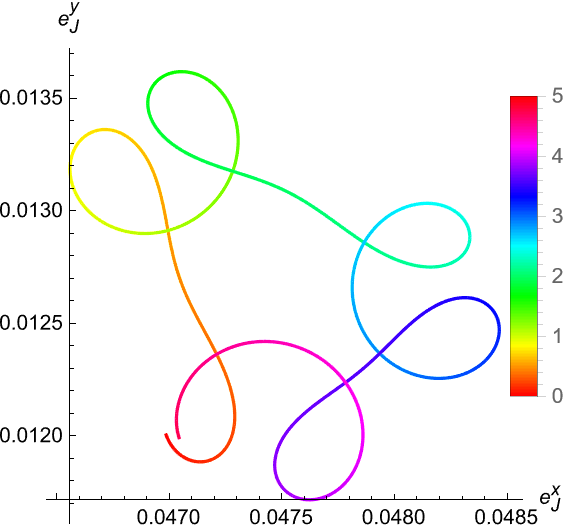}}
	\subfigure[\label{ecc2}]{\includegraphics[scale=0.5]{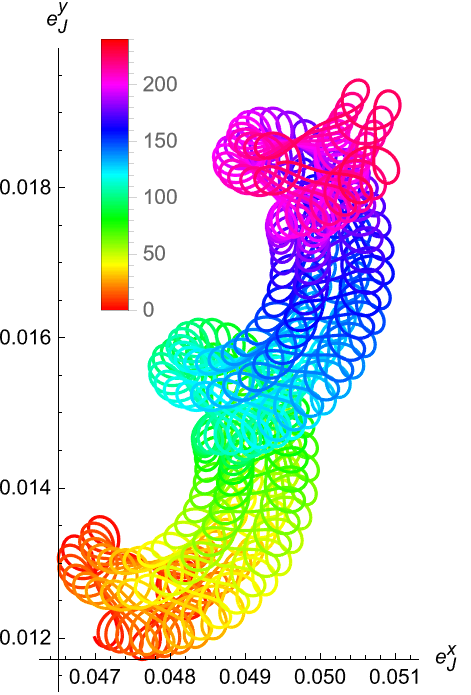}}
	\subfigure[\label{ecc3}]{\includegraphics[scale=0.55]{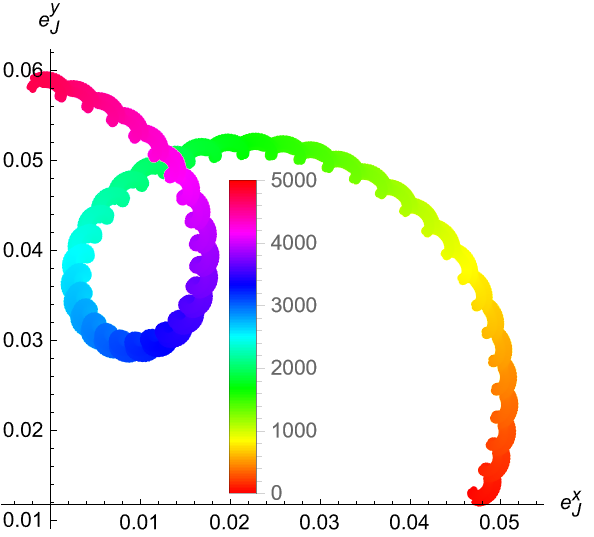}}
	\subfigure[\label{ecc4}]{\includegraphics[scale=0.4]{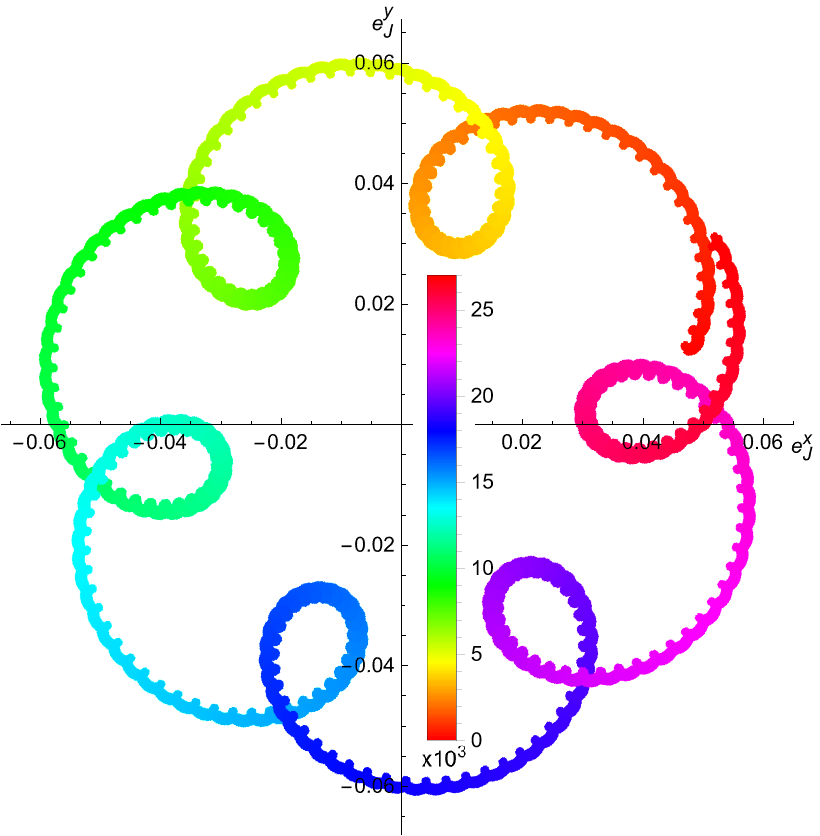}}
	\caption{Numerical solutions of the system (\ref{fs1}-\ref{fs4}) for Jupiter's eccentricity vector \\ $\vec{e}_1=(f_1^r(t)-1)\hat{r}(\theta_1(t))+f_1^\theta(t)\hat{\theta}(\theta_i(t))$, over different time-scales, with initial conditions based on the J2000 Epoch\cite{elements}. \ref{ecc1} shows the solution for 5 Jupiter-years (Jyrs), which is the cycle of the Jupiter-Saturn conjunctions, and reveals a repeating triangular pattern.  \ref{ecc2} is for 240 Jyrs, showing that the small `scallops' of the final plot happen on the scale of every 100 Jyrs. \ref{ecc3} plots from $t=0$ to $t=5000$ and shows that the large `scallops' are on the order of 4500 Jyrs. \ref{ecc4} plots for 27 kJyrs, and reveals a hexagonal pattern; indeed, 27,000/4500=6.}
	\label{Jecc}
\end{figure}

\section{Other formulations}\label{others}

For completeness, we present the following forms of the equations for both $\dot{\vec{e}}_i$ and $\dot{\vec{f}}_i$.   First of all, for $\dot{\vec{e}}_i$ in the polar coordinates as above
	\begin{align}
		\dot{e}_i^r&=2\frac{\tau_i}{\ell_i}(1+e^r_i)+\alpha^2\frac{(1+e_i^r)^2}{\ell_i^3}e_i^\theta \label{ecomp1}\\
		\dot{e}_i^\theta&=-\left(\frac{\sigma_i}{\ell_i}+\alpha^2\frac{e_i^r(1+e_i^r)}{\ell_i^3}\right)(1+e_i^r)+\frac{\tau_i}{\ell_i}e_i^\theta\,.\label{ecomp2}
	\end{align}
\noindent Of course, these are related to (\ref{fnew1},\ref{fnew2}) as the components of $\vec{e}_i,\vec{f}_i$ are related by
	\begin{equation}
		f_i^r=1+e_i^r,\quad f_i^\theta=e_i^\theta.
	\end{equation}
Another option, using either the eccentricity or auxiliary vectors, is to express them in their own polar coordinates; that is,
	\begin{align}
		\vec{e}_i(t)&=e_i(t)\hat{r}(\beta_i(t)) \\
		\vec{f}_i(t)&=f_i(t)\hat{r}(\psi_i(t)).
	\end{align}
\noindent Here, $e_i$ are the eccentricities (of the osculating orbits) and $\beta_i$ are the longitudes of periapsis, whereas $f_i$ are the lengths of the auxiliary vectors and $\psi_i$ are the angles which $\vec{f}_i$ make with the fixed reference (positive $x$-axis, for instance).  These definitions give
	\begin{align}
		\dot{\vec{e}}_i=\dot{e}_i\,\hat{r}_{\beta_i}+e_i\,\dot{\beta}_i\,\hat{\theta}_{\beta_i} \\
		\dot{\vec{f}}_i=\dot{f}_i\,\hat{r}_{\psi_i}+f_i\,\dot{\psi_i}\,\hat{\theta}_{\psi_i}
	\end{align}
\noindent where here the notation $\hat{r}_\times$ has been used for $\hat{r}(\times(t))$.  From these, we derive differential equations for norms ($e_i,f_i$) and arguments ($\beta_i,\psi_i$) by taking $\hat{r}_{\beta_i},\hat{\theta}_{\beta_i}$-components\footnote{These are derived with the following helpful properties of the $\hat{r},\hat{\theta}$ functions:  $\hat{r}(a)\cdot\hat{r}(b)=\hat{\theta}(a)\cdot\hat{\theta}(b)=\cos(a-b)$, $\hat{r}(a)\cdot\hat{\theta}(b)=\sin(a-b)$.} of (\ref{ecc.vel_sub}) and $\hat{r}_{\psi_i},\hat{\theta}_{\psi_i}$-components of (\ref{f_de_sub})
	\begin{align}
		\dot{e}_i&=\hat{r}_{\beta_i}\cdot\dot{\vec{e}}_i=\frac{\tau_i}{\ell_i}\left[e_i+(1+f_i^r)\cos(\theta_i-\beta_i)\right]+\frac{\sigma_i}{\ell_i}f_i^r\sin(\theta_i-\beta_i) \label{ebeta1} \\
		e_i\,\dot{\beta}_i&=\hat{\theta}_{\beta_i}\cdot\dot{\vec{e}}_i
		= \frac{\tau_i}{\ell_i}(1+f_i^r)\sin(\theta_i-\beta_i) -\frac{\sigma_i}{\ell_i}f_i^r\cos(\theta_i-\beta_i)  \label{ebeta2}
	\end{align}
\noindent while the equations for $f_i,\psi_i$ are
	\begin{align}
		\dot{f}_i=\hat{r}_{\psi_i}\cdot\dot{\vec{f}}_i
		&=\frac{\tau_i}{\ell_i}\left(f_i+f_i^r\cos(\psi_i-\theta_i)\right)+\frac{f_i^r}{\ell_i}\left(\alpha^2\frac{f_i^r}{\ell_i^2}-\sigma_i\right)\sin(\psi_i-\theta_i) \label{fpsi1}\\
		f_i\,\dot{\psi}_i=\hat{\theta}_{\psi_i}\cdot\dot{\vec{f}}_i 
		&= -\frac{\tau_if_i^r}{\ell_i}\sin(\psi_i-\theta_i) + \frac{f_i^r}{\ell_i} \left(\alpha^2\frac{f_i^r}{\ell_i^2} - \sigma_i\right) \cos(\psi_i-\theta_i) \label{fpsi2}.
	\end{align}
The systems $\lbrace \dot{\theta}_i,\dot{\ell}_i,\dot{e}_i,\dot{\beta}_i \rbrace$ and $\lbrace \dot{\theta}_i,\dot{\ell}_i,\dot{f}_i,\dot{\psi}_i \rbrace$ can be closed by the relations
	\begin{equation}
		f_i^r=f_i\cos(\psi_i-\theta_i)=1+e_i^r=1+e_i\cos(\theta_i-\beta_i).
	\end{equation}

\part{Perturbation Theory for the Three Body Problem}\label{part2}

\section{Delaunay Variables - an interpretation of the Mean Anomaly}\label{delaunay}

We will not here reproduce the whole definition and derivation of action-angle variables for the two-body problem, nor the canonical transformation to what are known as the Delaunay variables.  Suffice it now to say, that the Delaunay variables are a set of action-angle variables for the two-body problem, and that using these variables the Hamiltonian depends on none of the conjugate coordinates, revealing that the Hamiltonian is integrable, and the dynamics incredibly simple.  Excellent references for the material can be found in \cite{celletti, morbidelli}.

The two-body problem, of masses $m$ and $M$ orbiting each other subject to an attractive inverse-square-of-distance force $\vec{F}=-k/r^3\vec{r}$, where $\vec{r}$ is the relative position of one body to the other, reduces to a ``1-body" problem of a mass $\mu=mM/(m+M)$, called the \textit{reduced mass}, moving about a fixed centre by the same inverse-square force, and the displacement of the reduced mass from the centre is equal to the relative separation of the original bodies.  This problem has the equation of motion $\mu\ddot{\vec{r}}=-k/\abs{\vec{r}}^3\vec{r}$, and this may be given by the Lagrangian $\mathcal{L}=\tfrac{1}{2}\,\mu|\dot{\vec{r}}|^2+k/\abs{\vec{r}}$; equivalently the Hamiltonian $\mathcal{H}=|\vec{p}|^2/(2\mu)-k/\abs{\vec{r}}$, where the linear momentum is $\vec{p}=\nabla_{\dot{\vec{r}}}\mathcal{L}=\mu\dot{\vec{r}}$.  The only parameter essential to the problem is the ratio $\alpha=k/\mu$, and in the case of the gravitational two-body problem this is $GmM/\mu=G(M+m)$, called the standard gravitation parameter (sometimes `standard' is dropped).

Many things are well known of this problem and it's solutions.  It is straight forward to confirm that the angular momentum $\vec{L}=\mu\,\vec{r}\times\vec{v}$ is conserved, and that consequently the motion is planar.  It is also well known that the orbits $\lbrace\vec{r}(t)|\,t\in T\rbrace$ are conic sections, including ellipses for negative energies $E=\mu\,v^2/2-k/r$ (with the fixed `centre' at one of the foci of the ellipse), which will be our focus.  The physical size of elliptical orbits is characterized by the semi-major axis $a$, and it is also well known that the period of elliptical trajectories is given vy Kepler's Third Law: the square of an orbital period is proportional to the cube of the length of the semi-major axis
	\begin{equation}
		T^{-2}a^3=\alpha/4\pi^2
	\end{equation}
\noindent or equivalently, $a^3\omega^2=\alpha$, where $\omega=2\pi/T=\sqrt{\alpha/a^3}$, the angular frequency associated with period $T$, is called the \textit{mean motion}.

Before describing the Delaunay variables, we introduce various `specific' quantities---quantities that are made `massless', dividing by the chosen mass scale.  Thus the specific energy $\xi=v^2/2-\alpha/r$, and specific angular momentum $\vec{\ell}=\vec{r}\times\vec{v}$, $\ell=|\vec{\ell}|$.  These are both constants of the motion, given by $\xi=-\alpha/(2a)$ and $\ell=\sqrt{\alpha a(1-e^2)}$.
	\begin{figure}[h!]
		\includegraphics[scale=0.55]{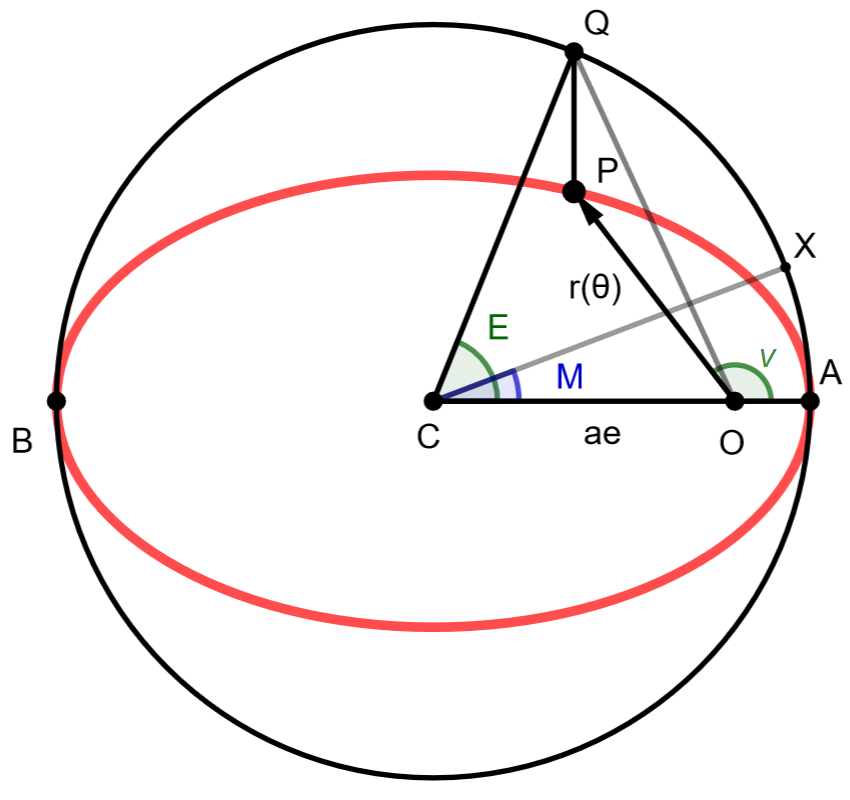}
		\caption{The construction of the eccentric and mean anomalies $E,M$.  $E=\angle ACQ$, while $M$ is defined such the area $\abs{AOQ}$ bounded by the circle is equal to the sector $\abs{ACX}$.}
		\label{geo}
	\end{figure}

With respect to a fixed orthonormal frame $\lbrace \hat{x},\hat{y},\hat{z}\rbrace$, the Delaunay momenta are: the $z$-component of angular momentum $\ell^z$, the total angular momentum $\ell$, and finally a third action $j$, related to both the energy and semi-major axis, given as $j=\sqrt{\alpha a}=\alpha/\sqrt{-2\xi}$.  For lack of another name, I will refer to this as the \text{impulse} of the orbit.  The coordinates are as follows\footnote{It should be noted that the Delaunay variables completely describe the 3D orientation of the elliptical orbit, as well as the progression of motion within the orbit.  The coordinates $\Phi$ and $\eta$ describe the axis of rotation of the orbital plane relative to the reference $xy$-plane and the position of periapsis within the orbital plane.  From the momenta $\ell^z,\ell,j$ can be determined $a,e$ and $\alpha$---in other words, the size, shape and speed of motion within the orbit.  The actual rotation of the orbital plane seems to be missing; that is, the inclination $\zeta$.  But since the angular momentum is perpendicular to the orbital plane, we can determine the inclination by $\cos(\zeta)=\ell^z/\ell$.}.  Conjugate to $\ell^z$ is the \textit{longitude of ascending node}, indicated with $\Phi$.  This is the angle from the positive $x$-axis to the ray along which the plane of the orbit intersects the $xy$-plane and where $\dot{z}>0$---this direction is called the \textit{ascending node}.  Conjugate to $\ell$ is the \textit{argument of periapsis}, $\eta$: the angle from the ascending node to the ray connecting the central body to the position of closest approach along the orbit, or periapsis, measured in the orbital plane.  The final angle, conjugate to the impulse $j$, is the so-called \textit{mean anomaly} $M$.  Unlike the previous two coordinates, $M$ is not so straightforward to define geometrically.  We must first construct the \textit{eccentric anomaly}, denoted $E$, by considering a circle of radius $a$ which circumscribes the ellipse, tangent at periapsis and apoapsis ($A$ and $B$ respectively).  The construction is shown in Fig.~\ref{geo}. The centre $C$ of this circle coincides with the centre of the ellipse.  We describe the position $P$ of the orbiting body around the ellipse by the angle subtended at the focus $O$ to periapsis, called the true anomaly and denoted $\nu$
	\begin{align}
		\vec{OP}=\vec{r}=r(\nu)&\hat{r}(\nu)\equiv r(\nu)\left(\cos\nu,\sin\nu\right) \\
		r(\nu)&=\frac{a(1-e^2)}{1+e\cos\nu}.
	\end{align}
\indent From $P$ we project to the auxiliary circle perpendicular to the major axis, arriving at a point $Q$.  The eccentric anomaly is the angle about $C$ from periapsis to $Q$, $E=\angle ACQ$.  Now, the ellipse can be recovered from the circle by a contraction of factor $(1-e^2)^{1/2}$ parallel to the minor axis.  This reveals that, working in the orbital plane and relative to the centre $C$, the position $P$ is
	\begin{equation}
		\vec{CP}=\left(a\cos E, b\sin E\right),
	\end{equation}
\noindent the axes being aligned with the major and minor axes, and $b=a\sqrt{1-e^2}$ is the semi-minor-axis.

Since the separation of the centre and focus is $CO=ae$, we have the relations
	\begin{align}
		\cos E&=\frac{(1-e^2)\cos\nu}{1+e\cos\nu}+e=\frac{\cos\nu-e}{1+e\cos\nu}\\
		\sin E&=\frac{\sqrt{1-e^2}\sin\nu}{1+e\cos\nu} \\
		\tan\tfrac{E}{2}&=\sqrt{\frac{1-e}{1+e}}\tan\tfrac{\nu}{2}
	\end{align}
\noindent and the inverse relations
	\begin{align}
		\cos\nu&=\frac{\cos E-e}{1-e\cos E} \\
		\sin\nu&=\frac{\sqrt{1-e^2}\sin E}{1-e\cos E}.
	\end{align}
\noindent These relations give the separation of the orbit as
	\begin{equation}
		OP=r(E)=a(1-e\cos E).
	\end{equation}
\indent We move to constructing the mean anomaly by employing Kepler's Second Law: we know that equal area is swept out by the orbit in equal times.  This can be expressed as the fact that the time-rate-of-change (t.r.o.c) of the area bounded by the ellipse $\mathcal{A}=\abs{AOP}$ is constant; in particular
	\begin{equation}
		\frac{d}{dt}(2\mathcal{A})=\ell.
	\end{equation}
\noindent Owing to the aforementioned contraction, the area bounded by the auxiliary circle $\tilde{\mathcal{A}}=\abs{AOQ}$ also grows uniformly with time
	\begin{equation}
		\frac{d}{dt}(2\tilde{\mathcal{A}})=(1-e^2)^{-1/2}\frac{d}{dt}(2\mathcal{A})=\sqrt{\alpha a}=j.
	\end{equation}
\noindent Thus, we may see the significance of $j$ as (twice) the t.r.o.c. of orbit-area projected to the auxiliary circle, just as angular momentum $\ell$ is twice the t.r.o.c. of area swept in the orbit itself.

The mean anomaly $M$ is defined as an angle in the auxiliary circle, measured at the center, say to a point $X$ on the circle, $M=\angle ACX$, such that the resulting sector has the same area as $\tilde{\mathcal{A}}=\abs{AOQ}$.  At a constant radius $a$, we see that the uniform growth of area $\tilde{\mathcal{A}}$ implies a constant rate-of-change for $M$
	\begin{align}
		2\abs{ACX}=a^2M\text{, so }a^2\dot{M}&=\frac{d}{dt}(2\tilde{\mathcal{A}})=\sqrt{\alpha a} \\
		\implies\quad \dot{M}&=\sqrt{\frac{\alpha}{a^3}}=\omega.
	\end{align}
\indent The mean and eccentric anomalies can be link by a straightforward calculation
	\begin{align}
		a^2M=2\abs{ACX}=2\abs{AOQ}&=2(\abs{ACQ}-\abs{OCQ}) \\
		&=a^2E-2(ae)(a\sin E)/2 \\
		\therefore\quad M&=E-e\sin E \label{ecc.to.mean}.
	\end{align}
\noindent We thus have the relation
	\begin{equation}
		E-e\sin E=\omega(t-\tau)
	\end{equation}
\noindent where $\tau$ is a time when the orbit is at periapsis.  A tantalizing equation, but unfortunately this cannot be inverted for the eccentric anomaly as elementary functions of time.  If this could be done, we could write the true anomaly explicitly as a function of the mean anomaly, and thus of time.  As it is, the relationship is implicit, although of course $\nu(M;e)$ is formally a well-defined function
	\begin{equation}
		\nu(M;e)=2\arctan\left(\sqrt{\frac{1+e}{1-e}}\tan\frac{E(M;e)}{2}\right);\quad 0\leq e<1
	\end{equation}
\noindent where $E(M;e)$ is the inverse\footnote{It can be seen that this inverse is a well-defined function as follows: for $\abs{e}\leq1$ the function $M(E;e)=E-e\sin E$ is increasing except for isolated points.  Where $M(E;e)$ is increasing, the inverse $E(M;e)$ is a differentiable function.  Even for $e=1$, on any domain $dM/de\geq0$, with equality only at $E=2k\pi,\,\,k\in\mathbb{Z}$. This gives $E(M;e)$ a continuous and increasing function, with $dE/dM\rightarrow+\infty$ for $M\in2\pi\mathbb{Z}$.} of (\ref{ecc.to.mean}) for given eccentricity $0\leq e\leq 1$.  We can see that (\ref{ecc.to.mean}) pairs $E=M=k\pi$ for $k\in\mathbb{Z}$.  If we consider $E(M;e)=M+\vartheta(M;e)$, this is $\vartheta=0$ for $M=k\pi$.  Furthermore, $\vartheta(M;e)$ is $2\pi$-periodic in $M$, and an odd function.  We can thus express $\vartheta$ as a $\sin$-Fourier series, with coefficients that are functions of $e$.
	\begin{equation}
		E(M;e)=M+\vartheta(M;e)=M+\sum_{k\geq1}c_k^{E}(e)\sin(kM)
	\end{equation}
\noindent These Fourier coefficients are $c_k^E(e)=2 J_k(ke)/k$, where $J_k(z)$ are the Bessel functions of the first kind.  We can see that the relationship $\nu(E)$ is just the same kind of relationship
	\begin{equation}
		\nu(E;e)=E+\sum_{k\geq1}c_k^\nu(e)\sin(kE).
	\end{equation}
\noindent Finally, it is straightforward to confirm that composition of such functions is closed---these functions being the sum of the identity function and some odd $2\pi$-periodic function.  We conclude that the expression for the true anomaly in terms of the mean anomaly has the same form
	\begin{equation}
		\nu(M;e)=M+\sum_{k\geq1}C_k(e)\sin(kM)
	\end{equation}
\noindent this Fourier sine-series is known as the ``equation of centre".  The coefficients have a remarkable expression in terms of Bessel functions
	\begin{equation}
		C_k(e)=\frac{2}{k}\left\lbrace J_k(ke)+\sum_{m\geq1}\beta^m\left[J_{k-m}(ke)+J_{k+m}(ke)\right]\right\rbrace,\quad\beta=\frac{1-\sqrt{1-e^2}}{e}\sim e/2+\cdots
	\end{equation}
\noindent These coefficients are of order $C_k(e)\sim \mathcal{O}(e^k)$; the Taylor series for the first of these functions begin
	\begin{align}
		C_1(e)&=2e-\frac{1}{4}e^3+\frac{5}{96}e^5+\frac{107}{4608}e^7+\cdots \\
		C_2(e)&=\frac{5}{4}e^2-\frac{11}{24}e^4+\frac{17}{192}e^6+\cdots \\
		C_3(e)&=\frac{13}{12}e^3-\frac{43}{64}e^5+\frac{95}{512}e^7+\cdots \\
		C_4(e)&=\frac{103}{96}e^4-\frac{451}{480}e^6+\cdots \\
		C_5(e)&=\frac{1097}{960}e^5-\frac{5957}{4608}e^7+\cdots \\
		C_6(e)&=\frac{1223}{960}e^6+\cdots \\
		C_7(e)&=\frac{47273}{32256}e^7+\cdots
	\end{align}

\subsection{Auxiliary Circular Orbit}

We now discuss two different proposals for what could be considered the ``auxiliary circular orbit", auxiliary to a given elliptical orbit.  The pedagogical idea is to be able to compare/connect the motion of a body in a elliptic Keplerian orbit, to a corresponding position in a related circular orbit.  This has the flavour of a ``method of images":  the (uniformly orbiting) position in the circular orbit is an image of the position in the elliptic orbit, the motion of which is non-uniform, and perhaps more difficult to develop an intuition for.  The relationship between the image-position and the elliptical position will ultimately be the geometrical one, based on the definition of the mean anomaly.  Indeed, we will take the mean anomaly, as determined in the elliptical orbit, to be the angular position of a body in a circular orbit about the same centre (the focus of the ellipse). 

Before proceeding, it should be pointed out that we will think primarily in the orbital plane, or equivalently in two dimensions.  In two dimensions $\Phi$ and $\eta$ themselves are undefined (better to say, undefinable), but their sum $\beta=\Phi+\eta$ is the well defined longitude of periapsis.  Delaunay variables in two dimensions are the conjugate pairs $\ell,\beta$ and $j,M$.

Certainly, the obvious option that first presents itself is to take the orbit with eccentricity $e^\prime=0$ and radius $a^\prime=a$.  With $j=\sqrt{\alpha a},\,\ell=\sqrt{\alpha a(1-e^2)}$, $\omega=\sqrt{\alpha/a^3}$ and $\xi=-\alpha/2a$, this choice gives $\ell^\prime=j^\prime=j$, $\omega^\prime=\omega$ and $\xi^\prime=\xi$.  It should be noted than in this scenario we are also taking the same gravitational parameter $\alpha^\prime=\alpha$, ie. the same total mass in the `phantom' circular two body set-up as in the original.  This certainly has simplicity to it's advantage, and that the phantom orbit has the same period as the original.

However, since the definition of $M$ is so linked to Kepler's Second Law and the uniform t.r.o.c.~of area-sweep, it would seem desirable to consider an auxiliary circular orbit which has the same angular momentum $\ell^\prime=\ell$ as the original, as well as having the same period, requiring $\omega^\prime=\omega$.  With $e^\prime=0$, these conditions require not only taking a circular radius different than the semi-major axis of the elliptical orbit, but also considering a change to the gravitational parameter $\alpha^\prime\neq\alpha$.  These conditions are the following
	\begin{align}
		\sqrt{\alpha^\prime\,a^\prime}&=\sqrt{\alpha a(1-e^2)},\quad\sqrt{\alpha^\prime/{a^\prime}^3}=\sqrt{\alpha/a^3} \\
		\intertext{the solution to which is}
		a^\prime&=a(1-e^2)^{1/4},\quad\alpha^\prime=\alpha(1-e^2)^{3/4}.
	\end{align}
\noindent With equal t.r.o.c. of area as well as equal orbital duration, it follows that the orbits enclose the same area: the elliptical area $\pi ab=\pi a^2\sqrt{1-e^2}=\pi {a^\prime}^2$ is equal to the area of circle with radius $a^\prime$.  We thus see the sense of this auxiliary circular orbit, as the Keplerian orbit with the same period and \textit{bounding the same area}.  In this sense we really can say the orbits are of the same size.

That we have to consider a modified gravitational parameter is to say that we must consider the auxiliary circular orbit as having a different total mass.  But this should not be problematic to us.  We do well to remember that the frequency of a Keplerian orbit depends principally on the semi-major axis and the total mass of the bodies in the orbit (we will consider Newton's constant $G$ as universal).  So to take a circular orbit with a radius different from a given semi-major axis, if we want an orbit with the same frequency, we must take a different total mass.  All this is to suggest that we should consider two orbital scenarios---in which one is both double the size in linear dimension and has 8 times the total mass than the other---as more similar to each other than orbits that have merely equal total mass or equal semi-major axis.

This consideration would also seem to elevate the stance of the angular momentum $\ell$ as somehow principal over the impulse $j$:  we have $j^\prime=\ell^\prime=\ell\neq j$, $\xi^\prime=\xi(1-e^2)^{1/2}$.  Indeed, we are seeing that the angular momentum is more characteristic of the orbit, as it corresponds to the actual area-sweeping rate in the physical orbit, and we preserve that in our `phantom' circular orbit, whereas $j$ is the projected area rate in the $a$-radius circle that we construct around the ellipse.  It is seen that this circle has far less to do with the actual physics than the above proposed auxiliary circular orbit.

\section{Alternate Mass parametrization}\label{mass}

We have freedom to take a different parametrization of the masses and the coefficients $\mu_{1,2,3}$.  First note, from the mass distribution of the three masses $m_J,m_S,M$, that any dimensionless ratios of masses depends on two independent mass-ratios.  We can form the following
	\begin{align}
		\beta_1=m_J&/M_\Sigma = 9.54\times 10^{-4},\quad\beta_2=m_S/M_\Sigma = 2.85\times10^{-4} \nonumber \\
		&\beta_3=M/M_\Sigma = 0.99876 = 1- 1.2384\times10^{-3} \nonumber \\
		&\text{which are constrained by }\beta_1+\beta_2+\beta_3=1 \nonumber.
	\end{align}
\noindent Our freedom is in the mass scale $\tilde{m}$ we take to divide the Lagrangian (\ref{lag}), giving the coefficients
	\begin{align}
		\mu_1=\frac{Mm_J}{M_\Sigma 	\tilde{m}},\quad\mu_2=\frac{Mm_S}{M_\Sigma\tilde{m}},\quad\mu_3=\frac{m_Jm_S}{M_\Sigma\tilde{m}}
	\end{align}
\noindent If we take $\tilde{m}=\lambda m_J$ for some $\lambda>0$ (originally we had $\lambda=M/M_\Sigma=\beta_3$), then these are
	\begin{align}
		\mu_1=\frac{\beta_3}{\lambda},\quad\mu_2=\frac{\beta_2\beta_3}{\beta_1\lambda},\quad\mu_3=\frac{\beta_2}{\lambda},\quad\delta=\frac{\beta_2\beta_3}{\lambda}
	\end{align}
\noindent We can choose $\lambda$ such that $\delta=1-\beta_3=\beta_1+\beta_2=(m_J+m_S)/M_\Sigma$, in which case $\mu_3=\delta/\beta_3=\delta/(1-\delta)$.  This is
	\begin{align}
		\lambda=\beta_2\beta_3\frac{M_\Sigma}{m_J+m_S}=\beta_3\frac{m_S}{m_J+m_S}
	\end{align}
\noindent and this gives
	\begin{equation}
		\mu_1=\frac{m_J+m_S}{m_S},\quad\mu_2=\frac{m_J+m_S}{m_J}
	\end{equation}
\noindent which are related by $1/\mu_1+1/\mu_2=1$.  In terms of the original parameter $\mu=m_S/m_J\sim 0.3$, these are $\mu_1=1+\mu^{-1}$ and $\mu_2=1+\mu$. For the masses of Jupiter and Saturn, these are $\mu_1=4.340=4\tfrac{1}{3}+6.452\times10^{-3}$ and $\mu_2=1.2994=1.3-5.796\times10^{-4}$.  This new mass scale is $\tilde{m}=\frac{m_S}{m_J+m_S}\frac{M}{M_\Sigma}m_J=\frac{Mm_Jm_S}{M_\Sigma(m_J+m_S)}$.  We will redefine $\epsilon\equiv\mu_3=\delta/(1-\delta)=(m_J+m_S)/M$.  Notice that $\delta$ is always $0\leq\delta\leq1$.  For the planetary regime, certainly we would say $m_J+m_s\leq M$, so that $\delta\leq1/2$ and $\epsilon\leq1$.

It is instructive to return to the coefficient matrix of the equations (\ref{eqs})
	\begin{align}
		A&=G\left(\begin{array}{ccc}
			-(M+m_J) & m_S & m_S \\ m_J & -(M+m_S) & m_J \\ M & M & -(m_J+m_S)
			\end{array}\right) \\
		&=\alpha\left(\begin{array}{ccc}
			\beta_2-1 & \beta_2 & \beta_2 \\ \beta_1 & \beta_1-1 & \beta_1 \\ \beta_3 & \beta_3 & \beta_3-1
		\end{array}\right) \\
		&=\alpha\left(\begin{array}{ccc}
			-1+\delta/\mu_1 & \delta/\mu_1 & \delta/\mu_1 \\ \delta/\mu_2 & -1+\delta/\mu_2 & \delta/\mu_2 \\ 1-\delta & 1-\delta & -\delta
		\end{array}\right)
	\end{align}

\section{Hamilton's Equations}\label{hamil}

We may now proceed to consider the asymptotic analysis of the three body problem, in the planetary case: when two bodies (the planets) orbit a third, and their masses are also much smaller than the third, but not so much that the gravitational attraction between them is negligible.  Using the Delaunay variables defined in the three sectors, we have the Hamiltonian
	\begin{align}
		\mathbf{H}_\lambda(J_i,h_i,h^z_i,M_i,\eta_i,\Phi_i,\vec{\lambda})&=\sum_{i=1}^3\left\lbrace\,- \mu_i^3 \frac{\alpha^2}{2J_i^2}- \lambda\cdot\vec{r}_i\right\rbrace \label{3BHam} \\
		&=\sum_{i=1}^3\left\lbrace\,- \mu_i^3 \frac{\alpha^2}{J_i^2}\right\rbrace-\vec{\lambda}\cdot\left(\vec{r}_1+\vec{r}_2+\vec{r}_3\right) \nonumber 
	\end{align}
\noindent where the Delaunay momenta are the relative variables (as opposed to specific)
	\begin{equation}
		J_i=\mu_i j_i\,,\quad h_i=\mu_i \ell_i\,,\quad h^z_i=\mu_i \ell^z_i\,.
	\end{equation}
\indent We notice that, with the exception of the impulses, the Delaunay variables enter into this Hamiltonian via the geometry, in the vector positions $\vec{r}_i$.  The vector position in the $i^\text{th}$ sector is
	\begin{equation}
		\vec{r}_i=\frac{J_i^2}{\alpha \mu_i^2}\left(1-e_i\cos{E_i}\right)\hat{r}_i
	\end{equation}
\noindent The eccentricity is given by $h_i^2/J_i^2=1-e_i^2$, and the direction vector $\hat{r}_i$ is derived by the following sequence of rotations: we start with the unit vector in a fixed reference, say  $\hat{x}$-direction.  This is then rotated ccw about the $z$-axis by both the true anomaly (related to the eccentric) and the argument of periapsis, $\nu+\eta$.  At this stage we can imagine we have an elliptic orbit with periapsis at argument $\eta$ and ascending node at the positive $x$-axis, but no inclination.  We need to rotate ccw about the $x$-axis by the inclination $\zeta_i$.  Note that the rotation by $\zeta_i$ is given by $\cos\zeta_i=h^z_i/h_i$ and $\sin\zeta_i=\sqrt{1-{h^z_i}^2/h_i^2}$. Finally, we rotate again about the $z$-axis, bringing the ascending node to longitude $\Phi_i$.  Thus $\hat{r}_i$ is
	\begin{equation}
		\hat{r}_i=R^z_{\Phi_i}R^x_{\zeta_i}R^z_{\nu_i+\eta_i}\left[1,0,0\right]^\text{T}\,.
	\end{equation}
\noindent Now, the components as a result of the rotation by true anomaly, are given in terms of the eccentric by
	\begin{equation}
		R^z_{\nu_i}\left[1,0,0\right]^\text{T}=\left[\frac{\cos E_i+e_i}{1-e_i\cos E_i},\frac{\sqrt{1-e_i^2}\sin E_i}{1-e_i\cos E_i},0\right]^\text{T}\,.
	\end{equation}
So we can see the $i^\text{th}$ position vector is
	\begin{equation}
		\vec{r}_i=\frac{J_i^2}{\alpha \mu_i^2}\,R^z_{\Phi_i}R^x_{\zeta_i}R^z_{\eta_i}\left[\cos E_i+e_i,\sqrt{1-e_i^2}\sin E_i,0\right] \label{geo1}
	\end{equation}
\noindent The components of the remaining rotation matrix are
	\begin{equation}
		R^z_\Phi R^x_\zeta R^z_\eta=\left(
		\begin{array}{ccc}
			\cos\Phi \cos\eta-\sin\Phi\cos\zeta\sin\eta & -\cos\Phi\sin\eta-\sin\Phi\cos\zeta\cos\eta & \sin\Phi\sin\zeta \\
			\sin\Phi\cos\eta+\cos\Phi\cos\zeta\sin\eta & \cos\Phi\cos\zeta\cos\eta-\sin\Phi\sin\eta & -\cos\Phi\sin\zeta \\
			\sin\zeta\sin\eta & \sin\zeta\cos\eta & \cos\zeta
		\end{array}
		\right) \label{geo2}
	\end{equation}
From the Hamiltonian (\ref{3BHam}), Hamilton's equation are
	\begin{align}
		\dot{M}_i&=\frac{\partial \mathbf{H}_\lambda}{\partial J_i}=\mu_i^3\alpha^2J_i^{-3}+\frac{\partial \mathbf{R}}{\partial J_i}& \dot{J}_i&=\mu_i\dot{j}_i=-\frac{\partial \mathbf{R}}{\partial M_i} \label{hamil1}\\
		\dot{\eta}_i&=\frac{\partial \mathbf{R}}{\partial h_i}& \dot{h}_i&=\mu_i\dot{h}_i=-\frac{\partial \mathbf{R}}{\partial \eta_i} \\
		\dot{\Phi}_i&=\frac{\partial \mathbf{R}}{\partial h^z_i}& \dot{h}^z_i&=\mu_i\dot{h}^z_i=-\frac{\partial \mathbf{R}}{\partial \Phi_i} \\
		0&=\nabla_{\vec{\lambda}}\mathbf{H}_\lambda=\vec{S}\label{hamil4}
	\end{align}
\noindent for $i=1,2,3$, where $\mathbf{R}=-\vec{\lambda}\cdot\vec{S}$ is the function by which the Hamiltonian is perturbed and $\vec{S}=\vec{r}_1+\vec{r}_2+\vec{r}_3$ is the constraint.  We know that when the equations are satisfied subject to the constraint, $\vec{\lambda}$ takes the values (\ref{mult}) as determined in the first section, and thus also determined by the geometry (\ref{geo1},\ref{geo2}). Now, we must be careful with the partial derivatives of the perturbation.  At first blush, we should regard both $\vec{\lambda}$ and $\vec{S}$ as functions of the coordinates.  If $\chi$ is one of the canonical variables, then the derivative of the perturbation with respect to $\chi$ is
	\begin{equation}
		-\frac{\partial \mathbf{R}}{\partial \chi_i}=\frac{\partial\vec{\lambda}}{\partial \chi}\cdot\vec{S} + \vec{\lambda}\cdot\frac{\partial\vec{S}}{\partial\chi} \label{part}
	\end{equation}
In particular, we compute the potential $\partial_\chi\vec{S}$ without regard to the constraint, but by the function form of the constraint $\vec{S}=\vec{r}_1+\vec{r}_2+\vec{r}_3$.  If $\chi=\chi_i$ is a variable of the $i^\text{th}$ sector, then this is $\partial_{\chi_i}\vec{r}_i$.  We might similarly determine the derivative $\partial_{\chi_i}\vec{\lambda}$ by combining (\ref{mult},\ref{geo1},\ref{geo2}), but at this point we may evaluate (\ref{part}) subject to constraint $\vec{S}=0$, so that the first term vanishes.  Thus
	\begin{equation}
		\frac{\partial \mathbf{R}}{\partial \chi_i} = -\vec{\lambda} \cdot \frac{\partial\vec{r}_i}{\partial \chi_i}.
	\end{equation}
For some of the derivatives we have the following (suppressing indices $i$), using $c_\zeta=\cos\zeta$ and $s_\zeta=\sin\zeta$,
	\begin{align}
		M&=(1-e\cos E)dE \nonumber \\
		\implies\partial_M&=(1-e\cos 	E)^{-1}\partial_E\qquad\qquad\qquad\qquad\qquad\,\,\,\,
	\end{align}
	\vspace*{-52pt}
	\begin{alignat}{11}
		\partial_J
		&=& &\partial_J\vert_e &\,\,+&& \frac{\partial e}{\partial J}\,&\partial_e  \nonumber \\
		&=& &\partial_J\vert_e &\,\,+&&\,\, \frac{1-e^2}{eJ}\,&\partial_e \\
		\partial_h
		&=& \,\frac{\partial e}{\partial h}&\,\partial_e & +&& \frac{\partial c_\zeta}{\partial h}\,&\partial_{c_\zeta} &\,\,+\,\,&& \frac{\partial s_\zeta}{\partial h}\,&\partial_{s_\zeta} \nonumber \\
		&=& \,-\frac{1-e^2}{eh}&\,\partial_e & -&& \frac{c_\zeta}{h}\,&\partial_{c_\zeta} &\,\,+\,\,&& \frac{c_\zeta^2}{s_\zeta h}\,&\partial_{s_\zeta} \\
		\partial_{h^z}
		&=& \frac{\partial c_\zeta}{\partial h^z}&\,\partial_{c_\zeta} &+&& \frac{\partial s_\zeta}{\partial h^z}\,&\partial_{s_\zeta} \nonumber \\
		&=& \frac{c_\zeta}{h^z}&\,\partial_{c_\zeta}& -&& \frac{c_\zeta^2}{s_\zeta h^z}\,&\partial_{s_\zeta}.
	\end{alignat}
\noindent These will be instrumental to formulating the KAM theory for this perturbational problem.

\section{Setup for KAM Theory}\label{kam}

KAM Theory
is a perturbative approach for nearly integrable Hamiltonians, which are a perturbation away from depending on only the momenta $\bar{J}\in\mathbb{R}^n$
	\begin{equation}
		\mathbf{H}(\bar{J},\bar{\theta};\delta)=\mathbf{H}_0(\bar{J})+\,\mathbf{\widehat{H}}(\bar{J},\bar{\theta};\delta)\sim \mathbf{H}_0(\bar{J})+\sum_{k\geq1}\delta^k\mathbf{H}_k(\bar{J},\bar{\theta}).
	\end{equation}
One seeks a nearly-identical canonical transformation $\mathcal{C}:(\bar{J},\bar{\theta})\mapsto(\tilde{J},\tilde{\theta})$, given by a generating function
	\begin{align}
		\mathbf{\Psi}(\tilde{J},\bar{\theta};\delta)&=\tilde{J}\cdot\bar{\theta}\,+\,\mathbf{\widehat{\Psi}}(\tilde{J},\bar{\theta};\delta)\nonumber \\
		&\sim\tilde{J}\cdot\bar{\theta}\,+\,\sum_{k\geq1}\delta^k\mathbf{\Psi}_k(\tilde{J},\bar{\theta}),
	\end{align}
which gives
	\begin{align}
		\bar{J}=\nabla_{\bar{\theta}}\mathbf{\Psi}&=\tilde{J}+\nabla_{\bar{\theta}}\mathbf{\widehat{\Psi}}(\tilde{J},\bar{\theta};\delta) \\
		&\sim \tilde{J}\,+\,\sum_{k\geq1}\delta^k\,\nabla_{\bar{\theta}}\mathbf{\Psi}_k \\
		\tilde{\theta}=\nabla_{\tilde{J}}\mathbf{\Psi}&=\bar{\theta}+\nabla_{\tilde{J}}\mathbf{\widehat{\Psi}}(\tilde{J},\bar{\theta};\delta) \\
		&\sim \bar{\theta}\,+\,\sum_{k\geq1}\delta^k\,\nabla_{\tilde{J}}\mathbf{\Psi}_k.
	\end{align}
The goal of this transformation is that the Hamiltonian, in terms of the new coordinates, depends only on the new momenta
	\begin{equation}
		\mathbf{H}\circ\mathcal{C}^{-1}\,(\tilde{J},\tilde{\theta};\delta)=\tilde{\mathbf{H}}(\tilde{J};\delta)
	\end{equation}
Practically speaking, if we only do this to so many terms, say truncating $\mathbf{\widehat{H}}=\sum_{k=1}^N\delta^k\mathbf{H}_k$ and $\mathbf{\widehat{\Psi}}=\sum_{k=1}^N\delta^k\mathbf{\Psi}_k$, then the transformed Hamiltonian only depends on new coordinates $\tilde{\theta}$ at the $N+1$ order in $\delta$
	\begin{equation}
		\tilde{\mathbf{H}}(\tilde{J},\tilde{\theta};\delta)=\tilde{\mathbf{H}}_N(\tilde{J};\delta)+\mathcal{O}(\delta^{N+1})(\tilde{J},\tilde{\theta};\delta).
	\end{equation}
\indent We will need to decompose the perturbations into the average over angle-variables $\bar{\theta}$
	\begin{align}
		\langle\mathbf{\widehat{H}}\rangle(\bar{J};\delta)&=(2\pi)^{-n}\oiint\mathbf{\widehat{H}}(\bar{J},\bar{\theta};\delta)\,\dd^n\bar{\theta} \\
		\langle\mathbf{H}\rangle_k(\bar{J};\delta)&=(2\pi)^{-n}\oiint\mathbf{H}_k(\bar{J},\bar{\theta};\delta)\,\dd^n\bar{\theta}
	\end{align}
and the remainders $\mathbf{\widehat{F}}=\mathbf{\widehat{H}}-\langle\mathbf{\widehat{H}}\rangle$, $\mathbf{F}_k=\mathbf{H}_k-\langle\mathbf{H}\rangle_k$.

Starting with just one order in $\delta$, $N=1$, writing $\bar{\mathbf{\Omega}}_0(\bar{J})=\nabla_{\bar{J}}\mathbf{H}_0$ for the frequency functions of the unperturbed Hamiltonian, we find
	\begin{align}
		\mathbf{H}(\bar{J},\bar{\theta};\delta)&\sim\mathbf{H}_0(\tilde{J}+\nabla_{\bar{\theta}}\mathbf{\widehat{\Psi}})+\delta\,\mathbf{H}_1(\tilde{J}+\nabla_{\bar{\theta}}\mathbf{\widehat{\Psi}},\bar{\theta})+\cdots \\
		&\sim\mathbf{H}_0(\tilde{J}+\delta\,\nabla_{\bar{\theta}}\mathbf{\Psi}_1+\cdots) 
		+\delta\,\langle\mathbf{H}\rangle_1(\tilde{J}+\delta\,\nabla_{\bar{\theta}}\mathbf{\Psi}_1+\cdots) \nonumber \\
		&\hphantom{=}\,\,+\delta\,\mathbf{F}_1(\tilde{J}+\delta\nabla_{\bar{\theta}}\mathbf{\Psi}_1+\cdots,\bar{\theta})+\cdots \\
		&\sim \mathbf{H}_0(\tilde{J})+\delta\,\bar{\mathbf{\Omega}}_0(\tilde{J})\cdot\nabla_{\bar{\theta}}\mathbf{\Psi}_1+\delta\,\langle\mathbf{H}\rangle_1(\tilde{J})+\delta\,\mathbf{F}_1(\tilde{J},\bar{\theta})+\mathcal{O}(\delta^2)
	\end{align} 
Thus we have
	\begin{equation}
		\tilde{\mathbf{H}}_1(\tilde{J};\delta)=\mathbf{H}_0(\tilde{J})+\delta\,\langle\mathbf{H}\rangle_1(\tilde{J}).
	\end{equation}
That is, the functional form of the transformed Hamiltonian, as a function of the new momenta, is the sum of the unperturbed Hamiltonian and the average of the perturbation over all angle variables, to leading order.  This is achieved by matching the remaining terms at first order
	\begin{equation}
		\bar{\mathbf{\Omega}}_0(\tilde{J})\cdot\nabla_{\bar{\theta}}\mathbf{\Psi}_1(\tilde{J},\bar{\theta})+\mathbf{F}_1(\tilde{J},\bar{\theta})=0 \label{kam1}
	\end{equation}
Expanding $\mathbf{\Psi}_1,\,\mathbf{F}_1$ in Fourier multi-series in $\bar{\theta}$, with coefficients $\Psi_1^{\bar{k}}(\tilde{J}),\,F_1^{\bar{k}}(\tilde{J})$, (\ref{kam1}) becomes (suppressing dependence on $\tilde{J}$)
	\begin{equation}
		\sum_{\bar{k}\in\mathbb{Z}^n\backslash\lbrace0\rbrace}\left\lbrace 
		\left(i\bar{\mathbf{\Omega}}_0\cdot\bar{k}\,\Psi_1^{\bar{k}}+F_1^{\bar{k}}\right)e^{i\bar{k}\cdot\bar{\theta}}
		\right\rbrace=0
	\end{equation}
So the solution is
	\begin{equation}
		\mathbf{\Psi}_1(\tilde{J},\bar{\theta})=i\sum_{\bar{k}\in\mathcal{S}_1} \frac{F_1^{\bar{k}}(\tilde{J})}{\bar{k}\cdot\bar{\mathbf{\Omega}}_0(\tilde{J})}\,e^{i\bar{k}\cdot\bar{\theta}}\label{kamsol1}
	\end{equation}
where $\mathcal{S}_1\subset\mathbb{Z}^n$ is the set of multi-indices for which $\mathbf{F}_1$ has non-zero Fourier coefficient.  Here we finally see a problem:  that if ever the frequency vector is orthogonal to one of these integer multi-indices $\bar{k}\cdot\bar{\mathbf{\Omega}}_0=0$, then the solution (\ref{kamsol1}) breaks down.  This is the problem of resonance, and it requires a modification called \textit{resonant perturbation theory}.  We have a special case of this problem: for our Hamiltonian (\ref{3BHam}), the unperturbed terms
	\begin{equation}
		-\mu_1^3\frac{\alpha^2}{J_1^2}-\mu_2^3\frac{\alpha^2}{J_2^2}\label{h0}
	\end{equation}
do not depend on the third impulse $J_3$ nor any of the angular momenta, which only come into the Hamiltonian through the perturbing function $\mathbf{R}$.  This results in the corresponding components of frequency vector being identically zero.  In other words, our Hamiltonian is degenerate, and the solution is called \textit{degenerate perturbation theory}.  What we need to do, not only decomposing the perturbation into it's average-over-all-angles and remainder, but further decomposing the remainder into it's average over the mean anomalies---the angles whose momenta the $0$th order Hamiltonian depends on (it will be seen that $M_3$ can be included, and the integrable term for the third sector can be included with the (\ref{h0}))---and remainder from that.  If $\mathbf{R}=-\vec{\lambda}\cdot\vec{S}=\langle \mathbf{R}\rangle+\mathbf{F}$ 
	\begin{align}
		&\langle\mathbf{R}\rangle(J_i,h_i,h^z_i)=(2\pi)^{-9}\oiint\,\mathbf{R}\,\,\dd^3 M\,\dd^3\eta\,\dd^3\Phi \\
		&\mathbf{F}(J_i,h_i,h^z_i,M_i,\eta_i,\Phi_i)=\mathbf{R}-\langle\mathbf{R}\rangle.
	\end{align}
Then we further decompose $\mathbf{F}= \mathbf{F}_M+\tilde{\mathbf{F}}$
	\begin{align}
		&\mathbf{F}_M(J_i,h_i,h^z_i,\eta_i,\Phi_i)=(2\pi)^{-3}\oiint\,\mathbf{F}\,\,\dd^3M \\
		&\tilde{\mathbf{F}}(J_i,h_i,h^z_i,M_i,\eta_i,\Phi_i)=\mathbf{F}_M-\mathbf{F}.
	\end{align}
So $\mathbf{R}=\langle \mathbf{R}\rangle+\mathbf{F}_M+\tilde{\mathbf{F}}$.
Then we seek a canonical transformation to a new Hamiltonian that doesn't depend on the mean anomalies.  This will be elaborated in future work.

\section{Conclusions}

In this work, we have presented a self-contained analysis of the gravitational three-body problem in the planetary scenario.  This is done in heliocentric coordinates with the use of Lagrange multipliers to elegantly handle the form of the Lagrangian.  First working in two dimensions, we develop the equations of motion as a first order dynamical system in longitudes $\theta_i$, angular momentum $\ell_i$ and eccentricity vectors $\vec{e}_i$.  Auxiliary vectors $\vec{f}_i=\hat{r}_i+\vec{e}_i$ greatly simplify the equations algebraically, especially in terms of the components $f_i^r,f_i^\theta$.

We then give the definition and construction of the Delaunay action-angle variables for the two body problem.  We present a novel conceptualization of the mean anomaly of an eccentric orbit as the true anomaly of a `phantom' or `image' body, also orbiting the central mass, but in a circular orbit with radius reduced from the semi-major axis $a^\prime=a(1-e^2)^{1/4}$ as well as reduced gravitational parameter $\alpha^\prime=\alpha(1-e^2)^{3/4}$.  These trajectories bound equal areas within the orbital plane, and the orbits have the same orbital frequency and angular momentum, as opposed to energy.  We then develop the Hamiltonian for the problem, and investigate the geometry of the orbital positions $\vec{r}_i$ in terms of the elements, in order to express the functional form of the Hamiltonian perturbation $\mathbf{R}=-\vec{\lambda}\cdot(\vec{r}_1+\vec{r}_2+\vec{r}_3)$.  Hamilton's equations are derived (\ref{hamil1}-\ref{hamil4}), and derivatives with respect to Delaunay variables are written down in terms of derivatives with respect to other elements $E,e,$ and $c_\zeta,s_\zeta$.  Finally, we considered the Hamiltonian perturbation theory of the problem, and learned that we need to proceed via a careful approach, seeing that our Hamiltonian is degenerate---not varying with all of the Delaunay momenta when $\delta=0$.

There is much room for further work.  The detailed work of the KAM theory analysis for this specific problem can now be begun.  Of particular interest is the Great Inequality of Jupiter and Saturn: to identify the term or terms that correspond to this perturbation that is unexpectedly large in both period and amplitude.  The equations in two dimensions, in terms of the auxiliary vectors, are interesting in their own right, and might be amenable to a multiple-scales asymptotic approach, as may be the Hamiltonian equations themselves, KAM theory aside.  Furthermore, the general-relativist corrections to this work are of great interest.  This would presumably be approached via the post-Newtonian formalism.


\begin{thebibliography}{99}
	\bibitem{principia} Newton, Isaac. Philosophiae Naturalis Principia Mathematica. 1687, London.
	
	\bibitem{laplace}	Laplace, Pierre-Simon. Théorie de Jupiter et de Saturne.  Memoire de l’Academie des Sciences de Paris 1788, 33-160.
	
	\bibitem{wilson}	Wilson, Curtis. The Great Inequality of Jupiter and Saturn: From Kepler to Laplace.  1985, Springer-Verlag.
	
	\bibitem{hill1}	Hill, G.W. Notes on the Theories of Jupiter and Saturn. The Analyst 1881 8(2), 33-40.
	
	\bibitem{hill2}	Hill, G.W. On the Extension of Delaunay’s Method in the Lunar Theory to the General Problem of Planetary Motion.  Trans. Amer. Math. Soc. 1900 1(2), 205-242.
	
	\bibitem{quarles}	Musielak, Z E and Quarles, B.  The Three-Body Problem. Rep. Prog. Phys. 2014 77 065901.
	
	\bibitem{kam}	Arnold, V.I. Proof of a Theorem by A.N. Kolmogorov on the invariance of quasi-periodic motions under small perturbations of the Hamiltonian.  Russ. Math. Survey 1963 18, 13-40.
	
	\bibitem{poincare}	Poincaré, Henri. Méthodes Nouvelles de la Mécanique Céleste, vol 1-3. 1892-99, Paris: Gauthier-Villars.
	
	\bibitem{chaos}	Feldman, David.  Chaos and Dynamical Systems. 2019, Princeton Univ. Press. ISBN:  9780691161525
	
	\bibitem{brouke_lass}	Brouke, R and Lass, H. A Note on Relative Motion in the General Three-Body Problem.  Celestial Mechanics 1973 8(1), 5-10.
	
	\bibitem{news1}	Hunt, Katie and Strickland, Ashley. “Jupiter and Saturn's 'great conjunction' captured in stunning images.” CTVNews.ca, Dec. 22, 2020. \url{https://www.ctvnews.ca/sci-tech/jupiter-and-saturn-s-great-conjunction-captured-in-stunning-images-1.5241665}
	
	\bibitem{news2}	Byrd, Deborah and McClure, Bruce. “All you need to know: 2020’s great conjunction of Jupiter and Saturn.”  EarthSky.org, Dec. 21, 2020. \url{https://earthsky.org/astronomy-essentials/great-jupiter-saturn-conjunction-dec-21-2020}
	
	\bibitem{celletti} Celletti, Alessandra.  Perturbation Theory in Celestial Mechanics. 2007, obtained from \url{https://web.ma.utexas.edu/mp_arc/c/07/07-303.pdf}
	
	\bibitem{morbidelli} Morbidelli, Alessandro. Modern Celestial Mechanics. 2011, obtained from \url{https://www-n.oca.eu/morby/celmech.pdf}
	
	\bibitem{elements} Seidelmann, K.P., ed. Explanatory Supplement to the Astronomical Almanac. 1992 University Science Books, Mill Valley, California.
\end{thebibliography}
\end{document}